\documentclass[apj]{emulateapj}

\slugcomment{}

\shorttitle{}
\shortauthors{}

\usepackage{color}
\usepackage{amsmath}
\usepackage{booktabs}

\usepackage{ulem}
\begin{document}


\title{IRAS\,16253--2429: the First Proto-Brown Dwarf Binary Candidate
Identified through Dynamics of Jets}

\author{Tien-Hao Hsieh$^{1, 2}$, Shih-Ping Lai$^{1}$, Arnaud Belloche$^{2}$ and Friedrich Wyrowski$^{2}$}
\affil{$^{1}$Institute of Astronomy, National Tsing-Hua University (NTHU), Hsinchu 30013, Taiwan}
\affil{$^{2}$Max-Planck-Institut f$\ddot{\textmd{u}}$r Radioastronomie (MPIfR), Bonn, Germany}

\email{shawinchone@gmail.com, slai@phys.nthu.edu.tw}

\begin{abstract}
%250 words as maximum
%Context
The formation mechanism of brown dwarfs (BDs) is one of the long-standing problems in star formation because the typical Jeans mass in molecular clouds is too large to form these substellar objects.
%Aim
To answer this question, it is crucial to study a BD at the embedded phase.
%Method
IRAS\,16253--2429 is classified as a very low luminosity object (VeLLO) with internal luminosity $<0.1L_\odot$.
VeLLOs are believed to be very low-mass protostars or even proto-BDs.
We observed the jet/outflow driven by IRAS\,16253--2429 in CO (2--1), (6--5), and (7--6) using the IRAM 30 m and APEX telescopes and the SMA in order to study its dynamical features and physical properties.
%Result
Our SMA map reveals two protostellar jets, indicating the existence of a proto-binary system as implied by the precessing jet detected in H$_2$ emission.
We detect a wiggling pattern in the position-velocity diagrams along the jet axes, which is likely due to the binary orbital motion.
Based on this, we derive the current mass of the binary as $\sim$0.032$M_{\odot}$.
Given the low envelope mass, IRAS\,16253--2429 will form a binary that probably consist of one or two BDs.
Furthermore, we found that the outflow force as well as the mass accretion rate are very low based on the multi-transition CO observations, which suggests that the final masses of the binary components are at the stellar/substellar boundary.
%Conclusion
Since IRAS\,16253 is located in an isolated environment, we suggest that BDs can form through fragmentation and collapse like low-mass stars.

\end{abstract}

{\let\thefootnote\relax\footnote{Based on observations carried out with the IRAM 30 m Telescope. IRAM is supported by INSU/CNRS (France), MPG (Germany) and IGN (Spain).}}
\section{INTRODUCTION}
\label{sec:int}
%VeLLOs and Proto Brown Dwarf
The formation mechanism of Brown Dwarfs (BDs, mass $<0.075 M_\odot$) is one of the highly debated unsolved problems because the typical Jeans mass in molecular clouds is too large to form these substellar mass objects \citep{pa04}. Three mechanisms have been proposed: 
%to answer this question: 
(1) In a molecular cloud, a fraction of cores with BD mass may be compressed by the
turbulent flow to reach sufficient high densities for the gravitational collapse to proceed \citep{pa04}. 
Oph-B11 was identified by \citet{an12} as a {\it pre-brown dwarf}, namely a starless core that will likely
form a BD in the future, suggesting this mechanism is a possible way to form a BD. 
This scenario is also supported by other works with identifications of BD candidates at early evolutionary stages \citep{ba09,pa14,mo15}.
(2) BDs are also considered to form in massive disks or multiple systems and be ejected later \citep{re01,ba02,ri03,st09,ba12}. (3) The third mechanism suggests that BDs form near massive stars which drive strong winds to disrupt the envelopes of BDs before they can accrete sufficient material to form stars \citep{wh04}. 
Identifying the formation mechanism(s) requires to study the early stage when a BD is still deeply embedded in its parental core, i.e., at the {\it ``proto-brown dwarf"} stage \citep{ba09}.

Discovered by the \textit{Spitzer} Space Telescope, Very Low Luminosity Objects (VeLLOs) are the faintest embedded protostars with internal luminosity $L_{\rm int}$ $<$ 0.1 $L_{\odot}$ \citep{di07}.
By comparing with evolutionary tracks from models, the low internal luminosity implies that VeLLOs remain substellar \citep{yo04,hu06,bo06} and are likely to form very low-mass stars or BDs in the future depending on their future accretion.
Several VeLLOs were considered to be proto-BD candidates in previous studies.
\citet{ba09} identified a proto-BD, J041757, in the Taurus molecular cloud based on 
the low luminosity and mass derived from its spectral energy distribution (SED).
The low mass accretion rate of L328-IRS calculated from its outflows suggests that L328-IRS would attain at most a mass of 0.05 $M_\odot$ \citep{le13}.
\citet{ka11} derived an unusually low density toward the natal core of L1148-IRS, implying it could be a good BD candidate.
\citet{pa14} found that IC 348-SMM2E will most likely remain substellar based on its low outflow force 
and low bolometric luminosity.
As a result, VeLLOs could be some of the best targets for studying BDs in the embedded phase.

%IRAS\,16253
IRAS\,16253--2429 (hereafter IRAS\,16253) was first discovered as a Class 0 source by \citet{kh04} in the $\rho$ Oph star forming region (d=125 pc, Evans et al.\ 2009). Later, \citet{du08} classified it as a VeLLO with an internal luminosity of $\sim$0.08--0.09 $L_{\odot}$.  IRAS\,16253 is located in a relatively isolated and quiescent portion in the east of the L1688 protocluster in the Ophiuchus molecular cloud complex. Thus it can be used for testing whether a BD can form in the same manner as a hydrogen-burning star.
\citet{to12} suggested that the central mass of IRAS\,16253 is less than 0.1 $M_\odot$ by studying the kinematic structure of the parent core through N$_2$H$^+$ observations with the Combined Array for Research in Millimeterwave Astronomy (CARMA).
\citet{ye15} also estimated the mass of the central star to be 0.02--0.04 $M_\odot$ using a kinematic model for a relatively small-scale C$^{18}$O emission using the Submillimeter Array (SMA).
In addition, IRAS\,16253 is believed to host a binary system based on the precessing H$_2$ jet detected by \citet{kh04} (see Section \ref{sec:pre}). This suggests that the mass of each component in the binary system is even lower.
To determine whether IRAS\,16253 is a proto-BD binary system or not, we need to measure the masses of the protostellar objects accurately and derive the mass accretion from the parent core.

%Why study outflows and jets??
It is difficult to determine the mass of the central star of a Young Stellar Object (YSO) especially at the embedded phase.
One can fit the SED to obtain the photospheric luminosity and the effective temperature of the protostar
and estimate the mass by comparing the value to evolutionary tracks \citep{hu06}.
%A popular approach of the central star mass is from the SED fitting.
However, one SED may be reproduced by several sets of parameters \citep{ro06, ro07}
and the resulting mass is model dependent. 
Recently, 
the central mass of YSOs has been obtained with high angular resolution interferometric observations from kinematic models
by assuming a Keplerian rotation in the disk or envelope \citep{to12,to12b,mu13,ye15}.
The Keplerian rotation is however difficult to detect in very low-mass objects like proto-BDs;
for example, both \citet{to12} and \citet{ye15} did not find significant velocity gradients in IRAS\,16253.
The most reliable mass estimation is resolving a binary rotation motion, which is mostly performed in pre-main sequence stars while the binary rotation is difficult to probe at the embedded phase.
In this study, we trace the binary orbital motion through the protostellar jets/outflows 
driven by IRAS\,16253 and derive the mass of the central stars in order to identify IRAS\,16253 as a proto-BD binary system.

Wiggling patterns of jets/outflows are used to probe the dynamics of binary systems.
This method has been applied toward several protostars
(IRAS 20126$+$4104, Shepherd et al.\ 2000; L1551, Wu et al.\ 2009; HH211, Lee et al.\ 2010; L1448C, Hirano et al.\ 2010; L1157, Kwon et al.\ 2015).
Wiggling patterns are believed to originate from (1) the orbital motion of the driving source or (2) the precession of the accretion disk caused by tidal interactions in a noncoplanar binary system. 
Both interpretations require the existence of a binary system, and the two origins can be distinguished by mirror-symmetric (orbital) or point-symmetric (precession) locus in both position-velocity (PV) diagrams and images \citep{ra09}.
The jet wiggling pattern caused by the orbital motion enables us to derive the orbital period, velocity, and binary separation, which allows us to determine the masses of the central stars.

In IRAS\,16253, a S-shaped (point-symmetric) H$_2$ jet was detected at 2.12 $\mu$m by \citet{kh04} and
in the mid-infrared by \citet{ba10}, but the spectral resolution was too low to probe the dynamics of the central stars.
\citet{st06} used the James Clerk Maxwell Telescope (JCMT) to map the large-scale CO (3--2) emission from the outflows.
However, the low-J CO emission is offset from the collimated H$_2$ jet, suggesting that they trace different gas.
In addition, the low-J transitions suffer from optical depth effects, resulting in an underestimate of the outflow mass as well as the outflow force \citep{du14,yi15}.
\citet{go13} further found that the molecular cloud and/or core emission could hide the outflows in low velocity regions, which makes the estimation more uncertain.

In this paper, we analyse multi-transition CO observations obtained with single-dish telescopes (CO 2--1 with the IRAM 30 m telescope, CO 6--5/7--6 with APEX), and an interferometer (CO 2--1 with the SMA).
We use the high spectral and spatial resolution data to study the dynamical structure of IRAS\,16253.
Mid-J CO (J = 6--5 and 7--6) emission traces the warm gas in protostellar outflows \citep{le09,ke09a,ke09b,ke09c,di09,yi12,go13,yi15} 
and is less affected by the above mentioned issues \citep{yi15}, which allows us to extract the outflow in low velocity regions and to accurately determine the outflow mass and force.

\begin{figure*}
%\vspace{2cm}
\includegraphics[scale=.470]{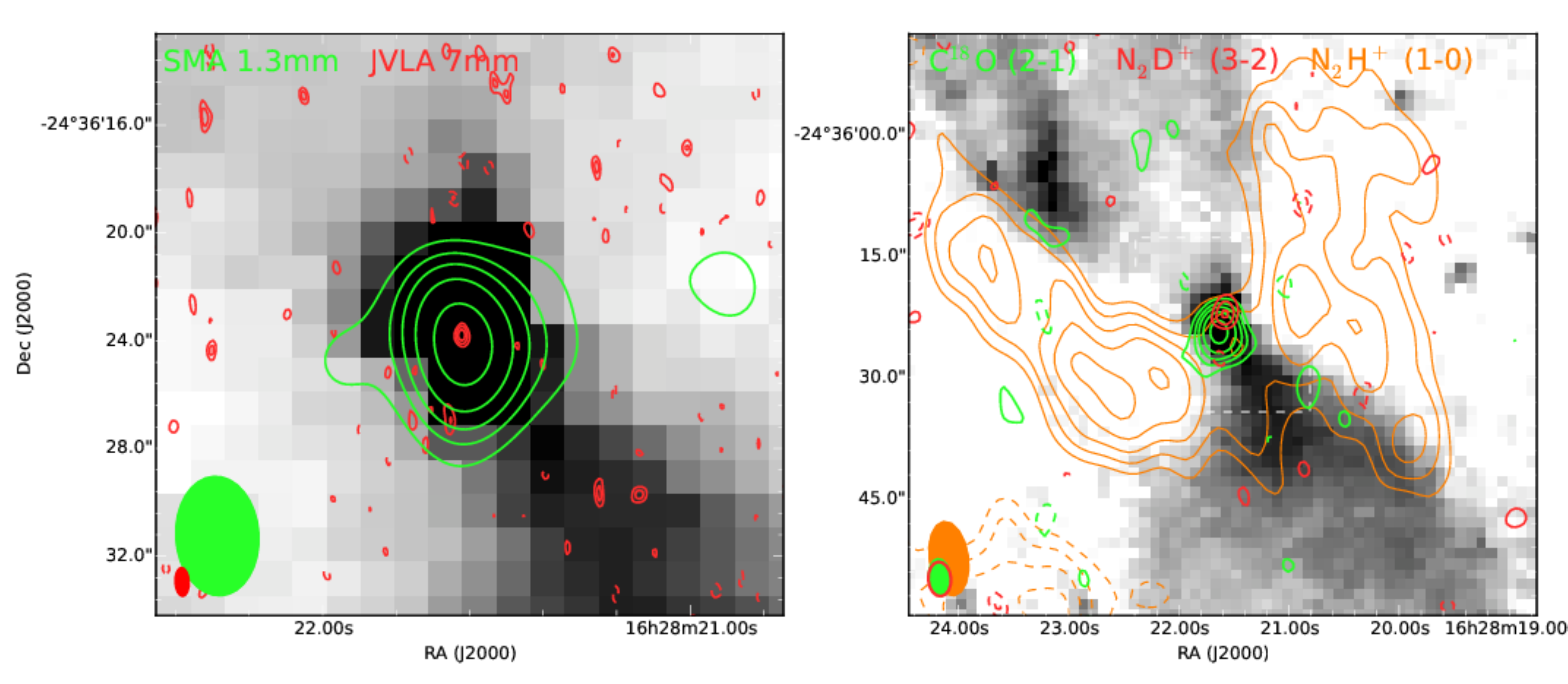}
\caption{(Left) SMA 224 GHz (green) and JVLA 43 GHz (red) continuum maps overlaid on an IRAC 1 image. The contour levels of the SMA map are 3, 5, 7 10, and 15$\sigma$ with a rms noise level, $\sigma = 0.87$ mJy beam$^{-1}$. The JVLA contours start at 3$\sigma$ with intervals of 1$\sigma$ in which $1\sigma$ is 0.02 mJy beam$^{-1}$.
(Right) SMA C$^{18}$O (2--1) (green), N$_2$D$^+$ (3--2) (red), and CARMA N$_2$H$^+$ (1--0) (orange) integrated intensity maps \citep{to11,to12} overlaid on the IRAC 1 image. The contour levels of C$^{18}$O are 3, 5, 7, 10, and 15$\sigma$ with a rms noise level, $\sigma = $64 mJy beam$^{-1}$ km s$^{-1}$. The N$_2$D$^+$ and N$_2$H$^+$ contours start at 3$\sigma$ and increase by 1$\sigma$ with $\sigma = 26$ mJy beam $^{-1}$ km s$^{-1}$ and $\sigma = 0.13$ Jy beam $^{-1}$ km s$^{-1}$, respectively. In both panels, the ellipses in the bottom left corner show the size of the beam, in the same color as the contours.}
\label{fig:sma}
\end{figure*}

\section{OBSERVATIONS}
\subsection{SMA observations}
%SMA
%CO 2-1, C18O 2-1, N2D+ 3-2
% track 1 and 2
We observed IRAS\,16253 using the Submillimeter Array (SMA) in the compact configuration in April 2013.
These observations targeted the 1.3 mm dust continuum emission and three molecular lines: CO (2--1) at 230.538 GHz, C$^{18}$O (2--1) at 219.560 GHz, and N$_2$D$^+$ (3--2) at 231.321 GHz.
The primary beam is about 55\arcsec~and the synthesized beam is $2\farcs8\times4\farcs0$ for CO (2--1). It is only slightly different for the other lines.
High spectral resolution windows were used for C$^{18}$O (2--1) and N$_2$D$^+$ (3--2) lines with 512 channels over 104 MHz, and for CO (2--1) with 256 channels over 104 MHz.
The resulting spectral resolutions are $\sim$0.28 km s$^{-1}$ for C$^{18}$O (2--1), $\sim$0.26 km s$^{-1}$ for N$_2$D$^+$ (3--2), and $\sim$0.53 km s$^{-1}$ for CO (2--1).
We used 3C279 as bandpass calibrator, J1626-298 as gain calibrator, and Neptune as flux calibrator ($\sim$10.5 Jy\footnote{http://sma1.sma.hawaii.edu}) for all observations.
The raw data were calibrated using the MIR package \citep{qi05} and the calibrated data were further imaged using MIRIAD \citep{sa95}.

The observations included (1) deep observations centered on the position of the infrared source 
($\alpha$=16h28m21.6s, $\delta$=$-$24\arcdeg36\arcmin23\farcs4, J2000)
in order to detect the dust continuum emission and weak molecular lines and 
(2) a mosaic for mapping the large-scale CO outflows toward three pointings, one at the northeast of the central
field and the other two at the southwest of the central field.  
The three pointing positions, combined with the central pointing position (four positions in total), are distributed along the outflow axis with an interval of $\sim$20\arcsec.
The deep observations contained one track using five antennas and one track using seven antennas. 
The on-source integration time was $\sim$5 hours for each track, and the sky opacity at 225 GHz was about 0.1.
These two data sets were later combined to increase the sensitivity.
For the mosaic mapping, one track was taken with also $\sim$5 hours of total on-source integration time for three positions using seven antennas.
The sky opacity at 225 GHz was about 0.08 during the observations.

\subsection{JVLA observations}
The Jansky Very Large Array (JVLA) observations were carried out 
in June 2013 toward the source center in the C configuration at 43 GHz. The total observing time was about 3 hours. The gain, bandpass, and flux calibrators were J1625-2527, J1256-0547, and J1331+3030, respectively. 
We used the Two 1-GHz mode with 16 subbands in total. 
The bandwidth of each subband was 128 MHz.
All but one subbands have a channel width of 1 MHz.
Only one subband was set to a higher spectral resolution of 31.3 kHz
for observing SiO (1--0) at 43.424 GHz.
The synthesized beam size is $1\farcs1\times0\farcs5$.
We used the calibrated visibilities processed through the VLA CASA (Common Astronomy Software Application)\footnote{http://casa.nrao.edu} calibration pipeline. The image in this paper was produced with CASA.

\subsection{IRAM 30 m CO (2--1) observations}
\label{sec:iramobs}
The CO (2--1) observations were carried out in June 2014 using the 30~m telescope of the Institute for Radio Astronomy in the Millimeter range (IRAM). 
The total observing time was about 9 hours and the sky opacity during the observations was 0.16--0.31.
The pointing and focus were checked every one to two hours and the pointing corrections were between 2\arcsec and 7\arcsec.
We used the E230 receiver and the FTS backend with a channel width of 50 kHz ($\sim$0.06 km s$^{-1}$). 
The data were later resampled to a channel width of 0.1 km s$^{-1}$. We used the on-the-fly mode to 
map a region of $84\arcsec\,\times\,228\arcsec$ with a position angle of 21\arcdeg~(east from north) and an angular resolution of $\sim$11\farcs2 at 230 GHz. The data were scaled from T$_{\rm a}^*$ to T$_{\rm MB}$ using a main beam efficiency of 0.58 and a forward efficiency of 0.92 taken from the IRAM 30~m website\footnote{http://www.iram.es/IRAMES/mainWiki/Iram30mEfficiencies}.
\subsection{APEX CO (6--5)/(7--6) observations}
\label{sec:apexobs}
We used the Carbon Heterodyne Array of the MPIfR (CHAMP$^+$) at the Atacama Pathfinder Experiment (APEX) \citep{gu08,ka06} to simultaneously observe CO (6--5) and (7--6) in May 2014. The total observing time was about 12.5 hours with a precipitable water vapor (PWV) in the range $\sim$0.4--0.7 mm.
The pointing and focus were checked every $\sim$1.5 hours and the pointing accuracy was found to be better than 1\arcsec.
The AFFTS backend was used with two overlapping 1.5 GHz units per pixel, resulting in 2.8 GHz band width and a spectral resolution of 212 kHz ($\sim$0.09 km s$^{-1}$ for 6--5 and 0.08 km s$^{-1}$ for 7--6).
The data were later resampled to a channel width of 0.1 km s$^{-1}$. 
The on-the-fly mode was used to map an area with a size of $80\arcsec\,\times\,300\arcsec$ with a position angle of 21\arcdeg~centered on the infrared source. 
The beam sizes are about 9\arcsec~for CO (6--5) and 8\arcsec~for CO (7--6) with a sampling every 3\arcsec. The data were smoothed to a resolution of 11\farcs2~in order to compare with the IRAM data.
The main beam efficiencies were set to 0.41 for CO (6--5) at 691.5 GHz and 0.34 for CO (7--6) at 806.7 GHz and the forward efficiency was set to 0.95 at both frequencies\footnote{http://www3.mpifr-bonn.mpg.de/div/submmtech/heterodyne/champplus/champ\_efficiencies.16-09-14.html}.

\begin{figure*}
%\vspace{2cm}
\includegraphics[scale=.58]{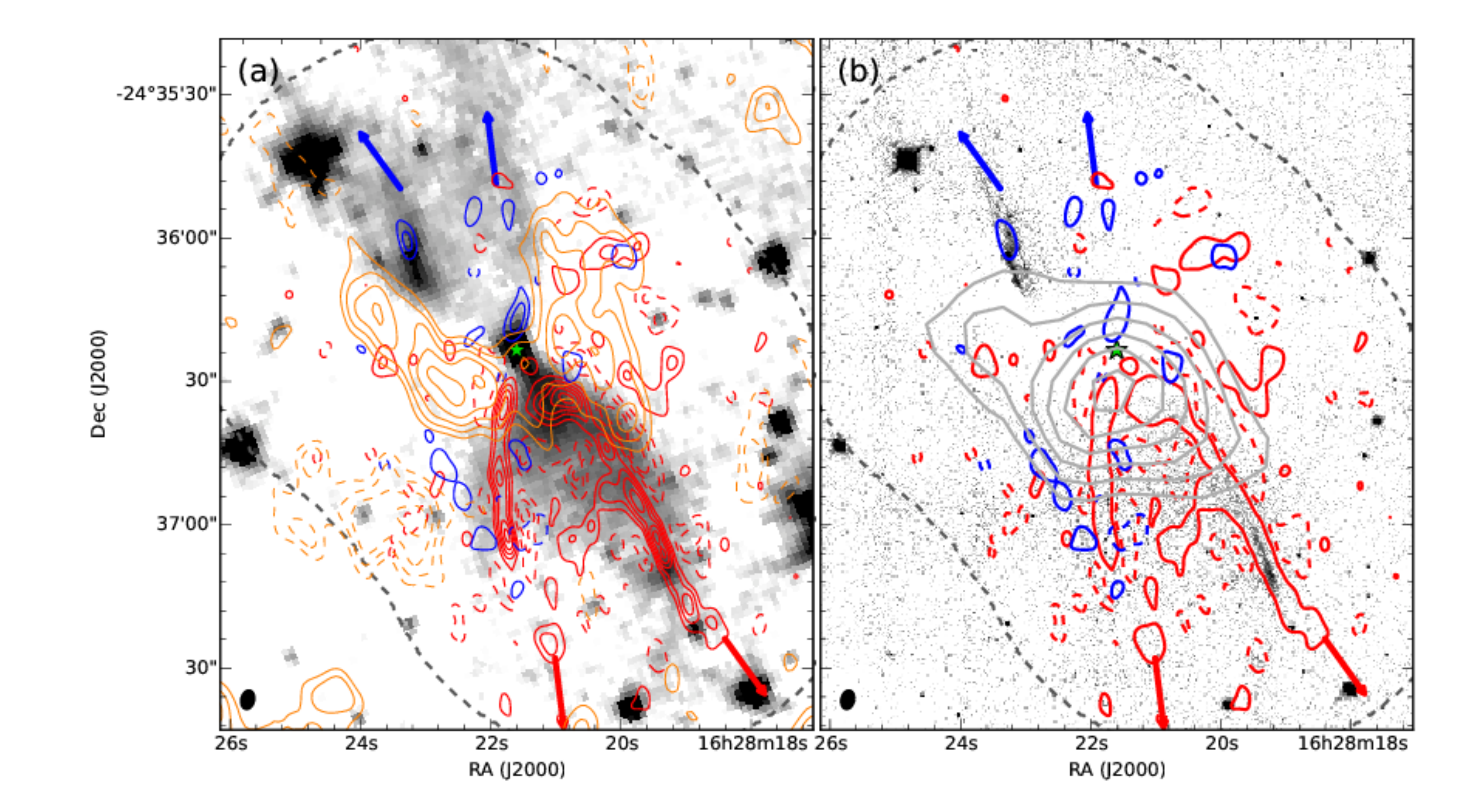}
\caption{(a) SMA CO (2--1) integrated intensity maps with velocities spanning 0.7--2.7 km s$^{-1}$ (blue) and 4.8--8.5 km s$^{-1}$ (red) overlaid on the \textit{Spitzer} IRAC 1 (3.6 $\mu$m) image. The contour levels are 5, 10, 15, 20, 25 and 30$\sigma$ with rms noise levels of $\sigma_{\rm blue} =$ 0.326 Jy beam$^{-1}$ km s$^{-1}$ and $\sigma_{\rm red} =$ 0.2 Jy beam$^{-1}$ km s$^{-1}$. 
The dashed line indicates the field of view of our SMA observations and the green star indicates the position of the infrared source.
The orange contours show the same N$_2$H$^+$ (1--0) map as Figure \ref{fig:sma}.
(b) Same as (a) but with only the 5$\sigma$ CO(2--1) contour overlaid on a CFHT H$_2$ image at 2.12 $\mu$m. The grey contours show the COMPLETE 850 $\mu$m continuum map with contour levels increasing from 3 to 8$\sigma$ in steps of 1$\sigma$, with $\sigma=0.056\,$Jy beam$^{-1}$.
}
\label{fig:smaco}
\end{figure*}

\section{RESULTS}
\subsection{Central source - SMA and JVLA images}
\label{sec:cen}
Figure \ref{fig:sma} (left panel) shows the continuum emission at 43 GHz from JVLA and 224 GHz from SMA.
They peak at the position of the infrared source, although the resolution of the two data sets is very different.
The dust emission at 224 GHz has a peak intensity of 16.7 mJy beam$^{-1}$. The source size is found to be $5\farcs0\times4\farcs6$ in $FWHM$ by fitting a 2-D Gaussian, and the deconvolved source size is $3\farcs4\times2\farcs3$.
The expected binary system, if it exists (see Section \ref{sec:pre}), is not resolved by the SMA observations with a synthesized beam size of $4\farcs0\times2\farcs8$ at a position angle of 3.6\arcdeg.
A point source was detected by JVLA at the position of the infrared source with a peak intensity of $\sim$0.12 mJy beam$^{-1}$.
The binary system is also not resolved by the JVLA observations with an angular resolution of $1\farcs1\times0\farcs5$, implying that the separation is extremely small.
However, another possibility is that the companion is too faint to detect at 43 GHz since the S/N ratio of the detected point source is only $\sim$5.9.

Figure \ref{fig:sma} (right panel) shows our C$^{18}$O (2--1) and N$_2$D$^+$ (3--2) integrated intensity maps.
The C$^{18}$O emission has a $FWHM$ size of $5\farcs7\times4\farcs7$ and a deconvolved size of $4\farcs2\times2\farcs8$.
It peaks at the position of the infrared source similar to the 224 GHz continuum emission, whereas the N$_2$D$^+$ (3--2) emission is point-like and peaks to the north of the infrared source at a distance of about $2\farcs6$ (325 au in projection).
The true distance could be larger than the projected distance if the C$^{18}$O and N$_2$D$^+$ emission comes from a flattened structure perpendicular to the outflow axis which is expected to lie close to the plane of the sky due to the collimated extended emission. 
The systemic velocities are found to be $4.01\pm0.04$ km s$^{-1}$ for the C$^{18}$O emission and $3.85\pm0.05$ km s$^{-1}$ for the N$_2$D$^+$ emission from hyperfine structure fitting,
although the fitting result of C$^{18}$O may be affected by its slightly asymmetric line profile that is skewed to the blue.

\begin{figure*}
%\vspace{2cm}
\includegraphics[scale=.53]{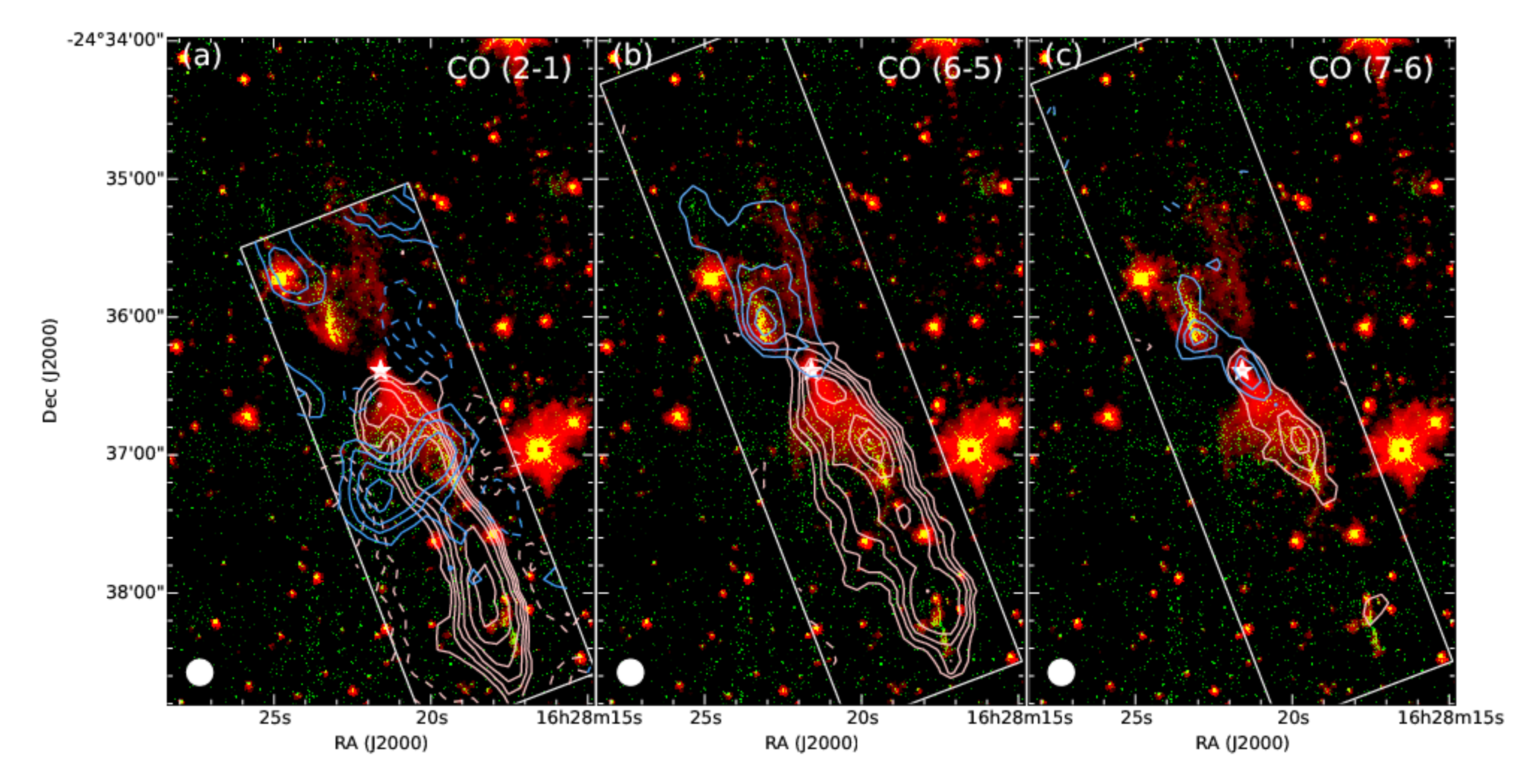}
\caption{(a) IRAM CO (2--1), (b) APEX CO (6--5), and (c) APEX CO (7--6) integrated intensity maps overlaid on a 2-color image with red scale for \textit{Spitzer} IRAC 1 (3.6 $\mu$m) and green scale for CFHT H$_2$ (2.12 $\mu$m).
The integrated velocity ranges are 0.65--2.75 km s$^{-1}$ (blue) and 4.75--8.55 km s$^{-1}$ (red) for the (2--1) data, 0.95--3.85 km s$^{-1}$ (blue) and 3.85--6.55 km s$^{-1}$ (red) for the (6--5) data, and 2.15--4.05 km s$^{-1}$ (blue) and 4.05--5.45 km s$^{-1}$ (red) for the (7--6) data. Note that the IRAM CO (2--1) map has the same velocity range as the SMA CO (2--1) map shown in Figure~\ref{fig:smaco}. 
The contour levels are -7, -5, -3, 3, 5, 7, 9, 11, 13, 15, and 20$\sigma$ for both IRAM and APEX data.
The rms noise levels are $\sigma_{\rm blue} =$ 0.66 K km s$^{-1}$ and $\sigma_{\rm red} =$ 0.61 K km s$^{-1}$ for the 2--1 data (a), $\sigma_{\rm blue} =$ 0.38 K km s$^{-1}$ and $\sigma_{\rm red} =$ 0.29 K km s$^{-1}$ for the 6--5 data (b), and $\sigma_{\rm blue} =$ 0.48 K km s$^{-1}$ and $\sigma_{\rm red} =$ 0.46 K km s$^{-1}$ for the 7--6 data (c).
The white star indicates the position of the infrared source and the white boxes show field of view of the contour maps.
}
\label{fig:sin}
\end{figure*}

\subsection{Jets/outflows from SMA observations}
\label{sec:smajet}
The SMA CO (2--1) integrated intensity map reveals two collimated components (Figure \ref{fig:smaco}).
Both components contain blue-shifted emission and red-shifted emission which are likely bipolar jets/outflows from the central submillimeter source.
One component lies along the North-East South-West direction (hereafter NE-SW), and the other is approximately along the North South direction (hereafter N-S) in the plane of the sky.
Each of them could trace either an outflow cavity wall or a jet.
Via scattered light, the outflow cavity is clearly seen in the IRAC 1 image (Figures \ref{fig:sma} and \ref{fig:smaco}) and exhibits a bipolar, symmetric, hourglass shape \citep{ba10}. 
The N$_2$H$^+$ (1--0) emission likely highlights the dense region carved by the bipolar outflows.
It is not seen toward the outflow cavity, especially on the side of the blue-shifted lobe. For the red-shifted lobe, the N$_2$H$^+$ emission partially overlaps with the cavity, which may be explained by projection effect; some N$_2$H$^+$ might be distributed in front of the outflow cavity.
The NE-SW CO (2--1) component is well separated from the edges of the outflow cavity traced by the near-infrared scattered light at 3.6 $\mu$m (Figure \ref{fig:smaco}a).
In addition, the NE-SW component matches well the H$_2$ emission (Figure \ref{fig:smaco}b, the H$_2$ image is obtained from the CFHT observations by T. Hsieh et al.\ 2016 in preparation) which exhibits a distinct S-shaped pattern and clearly does not follow the cavity edge \citep{ba10}; the CO emission continuously extends from the central source to the H$_2$ emission whereas the H$_2$ emission is detected in three isolated patches.
The southernmost H$_2$ patch in Figure \ref{fig:sin} does not have associated SMA CO emission, because it is outside the field of view of our SMA observations.
As a result, the NE-SW CO (2--1) component is most likely associated to the H$_2$ jet rather than to the wall of the outflow cavity.
The N-S CO (2--1) component is detected toward IRAS\,16253 for the first time. 
%It is unclear whether the N-S component traces an outflow cavity wall or a jet.
Like the NE-SW  component, the N-S component is not distributed along the edge of the outflow seen in the IRAC 1 image.
This suggests that the N-S component traces another jet, even if there is no H$_2$ emission associated with it.
The loci of both NE-SW and N-S components are curved and likely have a point symmetry at the source center. 
We hereafter call these two components NE-SW jet and N-S jet.
%We hereafter call these two components ``jets" (NE-SW jet and N-S jet), although we can not completely exclude the possibility that the N-S component is a cavity wall.
%In addition, the NE-SW component may also contain the contribution from the other side of the cavity wall.

\subsection{Jets/outflows from IRAM 30~m and APEX observations}
\label{sec:sin}
The CO (2--1), (6--5), and (7--6) integrated intensity maps are shown in Figure \ref{fig:sca}.
For a better comparison, the CO (6--5) and (7--6) maps were smoothed to the angular resolution of the CO (2--1) map (11\farcs2), and all the maps were constructed in the same grid with a pixel size of 5\arcsec.
Because the CO emission traces both the outflows and the large-scale structures of the molecular cloud, 
we applied a multiresolution analysis \citep{be11} to the channel maps in order to remove the cloud emission (see Section \ref{sec:sca}).
Figures \ref{fig:sin} and \ref{fig:sca} to \ref{fig:APEXch1bsmo} show the integrated intensity and channel maps after filtering the large-scale emission produced by the molecular cloud. % ABE: I don't understand this: and the emission being filtered out.

\begin{figure*}
%\vspace{2cm}
\includegraphics[scale=.52]{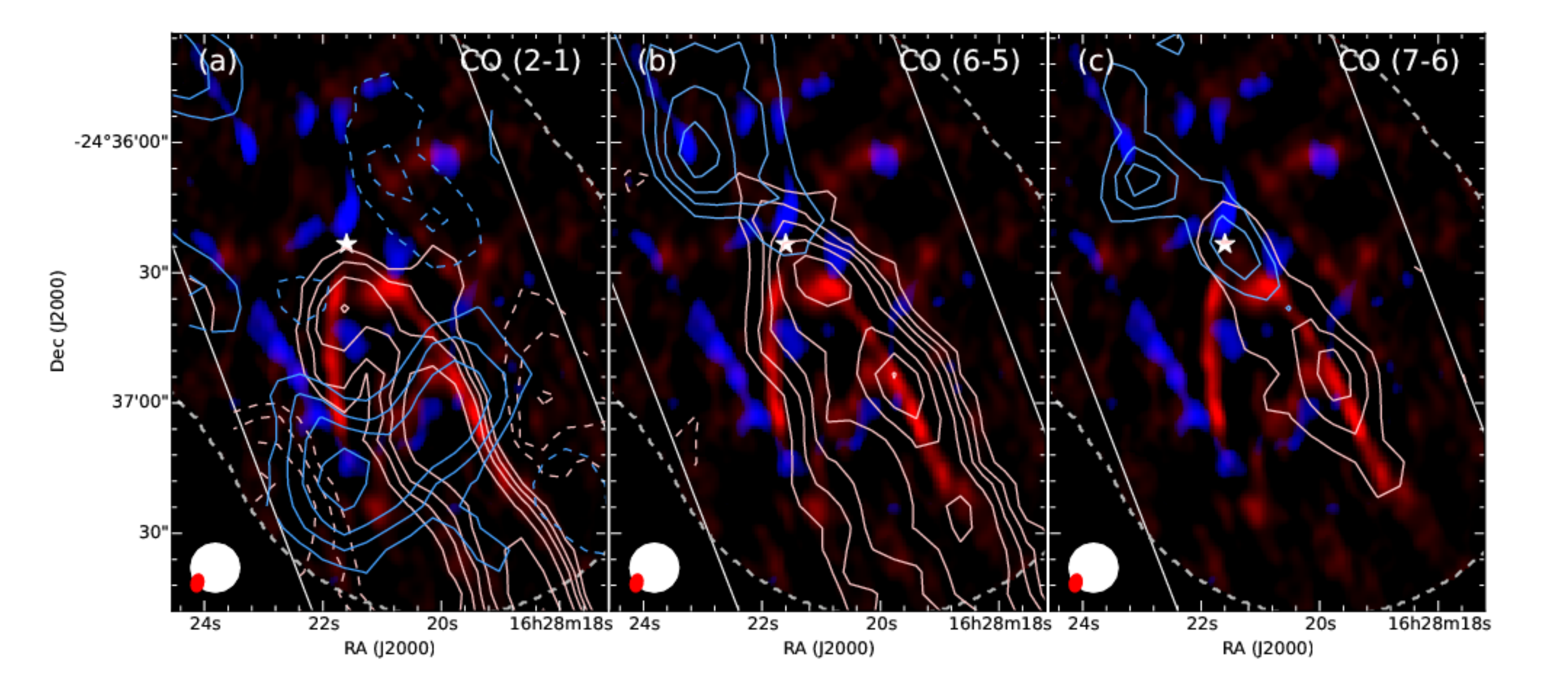}
\caption{Comparison of the SMA CO (2--1) integrated intensity map (color scale) and the single-dish maps (contour).
The maps are the same as in Figures~\ref{fig:smaco} and \ref{fig:sin}.
}
\label{fig:com}
\end{figure*}

%CO 2-1 and 6-5
The molecular outflows detected in CO (2--1) and (6--5) are extended up to $\sim$2\arcmin-3\arcmin\ (15000--23000~au) from the source center. 
For both CO (2--1) and (6--5), the blue-shifted outflows are much weaker than the red-shifted outflows.
However, the spatial and velocity distributions are very different between CO (2--1) and (6--5) as shown in the integrated intensity maps (Figure \ref{fig:sin}) and channel maps (Figures \ref{fig:IRAMch1} and \ref{fig:APEXch1}).
%(We note that the channel maps have go through a large-scale removed process, see section \ref{sec:sca})
%Velocity
The CO (6--5) emission tends to be stronger than CO (2--1) at low velocity;
CO (6--5) is much brighter in the velocity ranges of $\sim$2--3.5 km s$^{-1}$ (blue-shifted) and $\sim$4.4--6.4 km s$^{-1}$ (red-shifted), while CO (2--1) is brighter in the velocity ranges of $\sim$0.9--2.6 km s$^{-1}$ (blue-shifted) and $\sim$4.7--7.1 km s$^{-1}$ (red-shifted).
%Spatial
Besides, in the low velocity range of $\sim$4.7--5.0 km s$^{-1}$, CO (2--1) is undetected in the inner regions where the CO (6--5) emission is bright.
For the blue-shifted outflow lobe, the CO (6--5) emission is distributed from the driving source to its north-east at a distance of $\sim$100\arcsec, and CO (2--1) is only seen at a distance $\gtrsim 40''$ (Figure~\ref{fig:sin}).
The CO (2--1) blue-shifted outflow is likely extended over a larger region than our IRAM 30 m map which was designed to cover only the jet region detected in H$_2$.

%CO 7-6
CO (7--6) is detected in an even lower velocity range than CO (2--1) and (6--5) (Figures~\ref{fig:sin} and \ref{fig:APEXch1b}).
Spatially, CO (7--6) is mostly detected in the inner $1'$ 
region but the signal-to-noise ratio is low.
The CO (7--6) emission can be separated into two components. 
A compact component toward the center is seen in the velocity range {$\sim$3.3--4.1 km s$^{-1}$ for which there is no detection in CO (2--1) and (6--5).
An elongated component is detected in the velocity ranges of $\sim$2.3--2.9 km s$^{-1}$ (blue-shifted) and  $\sim$4.3--5.4 km s$^{-1}$ (red-shifted), which are associated to the collimated H$_2$ jet.

%Compare with H2
A particularly intriguing result is found when comparing the CO data with the shock tracer H$_2$ (Figure~\ref{fig:sin}).
We find that the CO (7--6) and (6--5) integrated intensities match the locus of the H$_2$ jet well, which has a ``S-shaped'' point symmetry around the driving source.
In contrast, the CO (2--1) integrated intensity is shifted to the east of the H$_2$ jets. It likely probes the outflow entrained gas and cavity wall detected in scattered light with IRAC 1.

%Compare with SMA
Figure~\ref{fig:com} shows a comparison of the SMA CO (2--1) integrated intensity map to the single-dish CO integrated intensity maps.
The red-shifted component of the N-S jet from the SMA map is also detected with the IRAM 30~m telescope. 
However, the SMA NE-SW jet associated with the H$_2$ jet is likely hidden by the large-scale outflow emission and/or cloud emission in the IRAM CO (2--1) map.
Nevertheless, the warm gas tracers CO (6--5)/(7--6) do trace the NE-SW jet very well
%and reveal it from the large-scale emissions
(Figures~\ref{fig:com}b and \ref{fig:com}c).
 
%\subsubsection{The southern blue-shifted component in CO (2--1)}
A blue-shifted emission is detected in the southern part in the CO (2--1) integrated intensity map (Figure \ref{fig:sin}a) and channel map (Figures \ref{fig:IRAMch1}).
This emission can also be seen in the IRAM and SMA CO (2--1) PV-diagrams (see Sections \ref{sec:pre} and \ref{sec:orbwig}).
The component extends outside the outflow cavity (Figure~\ref{fig:sin}) and its systemic velocity of $\sim$0.7--1.6 km s$^{-1}$ (Figure~\ref{fig:IRAMch1}) is relatively lower than the blue-shifted outflow in all single-dish and SMA maps.
Thus, this component is most likely not associated with the outflows.
%The nondetection in CO (6--5) could further support this argument because the whole cavity structures are clearly seen in the CO (6--5) maps (Figure \ref{fig:sinsca}).
Because its systemic velocity is also different from that of IRAS\,16253 as well as the Ophiuchus molecular cloud, we speculate that this blue-shifted emission originates from some background sources.

 \begin{deluxetable*}{cccccccc}
\tablewidth{0pt}
\tabletypesize{\tiny}
\tablecaption{Model Parameters of The Precessing Jets}
\tablehead{
\colhead{}    			&  \colhead{P.A. ($\psi$)}   & \colhead{$\alpha$} 	& \colhead{$\lambda_{\rm pre}$}	& \colhead{$\phi_{\rm 0,\, pre}$} & \colhead{$\theta_{\rm inc}$\tablenotemark{a}}	& \colhead{$\alpha'$\tablenotemark{b}}	& \colhead{$V_{\rm jet}$\tablenotemark{c}} \\
\colhead{Jet}      & \colhead{[deg]} 			& \colhead{}			& \colhead{[arcsec]}		& \colhead{[deg]} 		& \colhead{[deg]}	& \colhead{[deg]} &
\colhead{km s$^{-1}$}
}
\startdata
NE-SW				& 38.4$\pm$0.1					& 0.15$\pm$0.01			& 299$\pm$17		& 78.8$\pm$4.8			& 20 		& 8.5$\pm$0.4	 & 6.0 \\	
N-S					& 7.8$\pm$0.1			& 0.26$\pm$0.01			& 233$\pm$8				& 288.5$\pm$3.7			& 20		& 14.6$\pm$0.4	& -
\enddata
\tablecomments{Col.\ (1): Jet component. Col.\ (2): Position angle of the jet. Col.\ (3) Precession amplitude: Col.\ (4): Spatial period of the jet precession. Col.\ (5): Initial phase of precessing jet. Col.\ (6): Inclination angle of the jet. Col.\ (7): Half-opening angle of the precession cone. Col.\ (8) Jet velocity.}
\tablenotetext{a}{The inclination is adopted based on the morphology.}
\tablenotetext{b}{$\alpha'$ represents the angle between the precessing jet and its symmetry (precession) axis, which is obtained by $\alpha' = \arctan(\alpha)$.}
\tablenotetext{c}{Jet velocity estimated based on the PV diagram of the APEX data.}
\label{tab:prepa}
\end{deluxetable*}

\section{ANALYSIS}
\subsection{Dynamical features of the molecular jets/outflows}
\subsubsection{Filtering of the large scale emission}
\label{sec:sca}
%Arnaud 2011 method
%Remove large-scale component from cloud
Our single-dish CO (2--1)/(6--5)/(7--6) maps are severely contaminated by the large-scale cloud emission especially for the CO (2--1) transition.
To remove this extended emission, we applied a multiresolution analysis to the channel maps. 
This method (see Appendix C in Belloche et al. 2011 for details) is based on a median filter which extracts the structures at different scales of $<2^{\rm i}+1$ pixels where ${\rm i}=$1, 2, 3, 4, 5, and 6 with a pixel size of 5$''$.
The multiresolution analysis decomposes the initial map into maps containing small-scale structures (summation maps) and the corresponding maps containing large-scale structures (smooth maps). 
For example, in step 3 (${\rm i}=3$), the summation map includes structures at scales $<$ 9 pixels and the smooth map contains structures at scale $>$ 9 pixels; the addition of these two maps is strictly equal to the input (observed) map.
We checked the outcome of this decomposition at each step in order to select the step at which the
outflow emission and cloud emission are best separated.
As a result, we took step 4 for CO (2--1), step 5 for CO (6--5), and step 4 for CO (7--6).
We plot the integrated intensity summation maps, smooth maps, and input maps in Figure~\ref{fig:sca}.
The channel summation maps and channel smooth maps are shown in Figures \ref{fig:IRAMch1} to \ref{fig:APEXch1bsmo}.
In the following, we use the summation maps at the specific steps to derive the outflow properties.
The removal of large-scale emission appears to be crucial to derive the outflow properties,
especially for CO (2--1) for which the contribution of the cloud could even be a few times stronger than the contribution of the outflows.

\begin{figure}
\includegraphics[scale=.60]{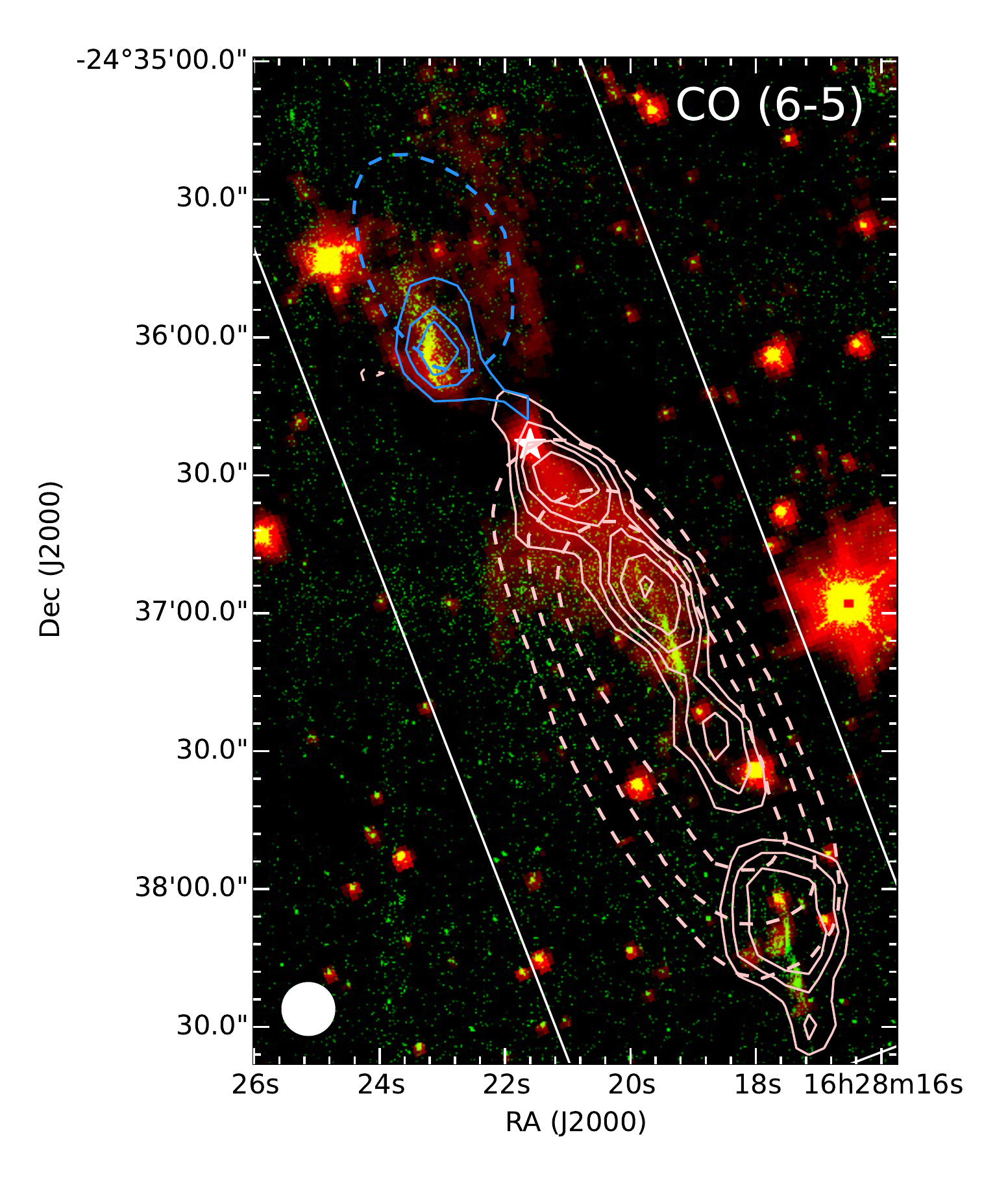}
\caption{CO (6--5) map of Figure~\ref{fig:sin} further decomposed into the step 3 summation map (solid line) and smooth map (dashed line) using a multiresolution analysis (see Section \ref{sec:sca}). 
The contour levels are 3, 5, 7, 10, 15$\sigma$ with $\sigma$ the same as in Figure~\ref{fig:sin}.
}
\label{fig:sinsca}
\end{figure}

%Wind or jet? is the very small scale trace collimated jet?
The multiresolution analysis is also used to extract the jet (knot) from the outflow cavity.
After removal of cloud emission, we further used the step 3 summation maps to separate the collimated NE-SW jet and the extended outflow cavity in CO (6--5) (Figure~\ref{fig:sinsca}).
We applied the multiresolution analysis to the step 5 summation map of CO (6--5) such that the sum of the two contour maps in Figure \ref{fig:sinsca} is equal to that in Figure \ref{fig:sin}b.
The integrated intensity summation map matches the H$_2$ jets well while the peak of its corresponding smooth map is shifted to the east like the CO (2--1) map;
the smooth map here includes only the extended structures in the outflows while the large-scale cloud emission has already been removed.
This result suggests that the small-scale emissions may trace the collimated jet and/or the jet knots.
In contrast, the large-scale structures are sensitive to the entrained gas and/or the outflow cavity wall.

\begin{figure*}
\includegraphics[scale=.65]{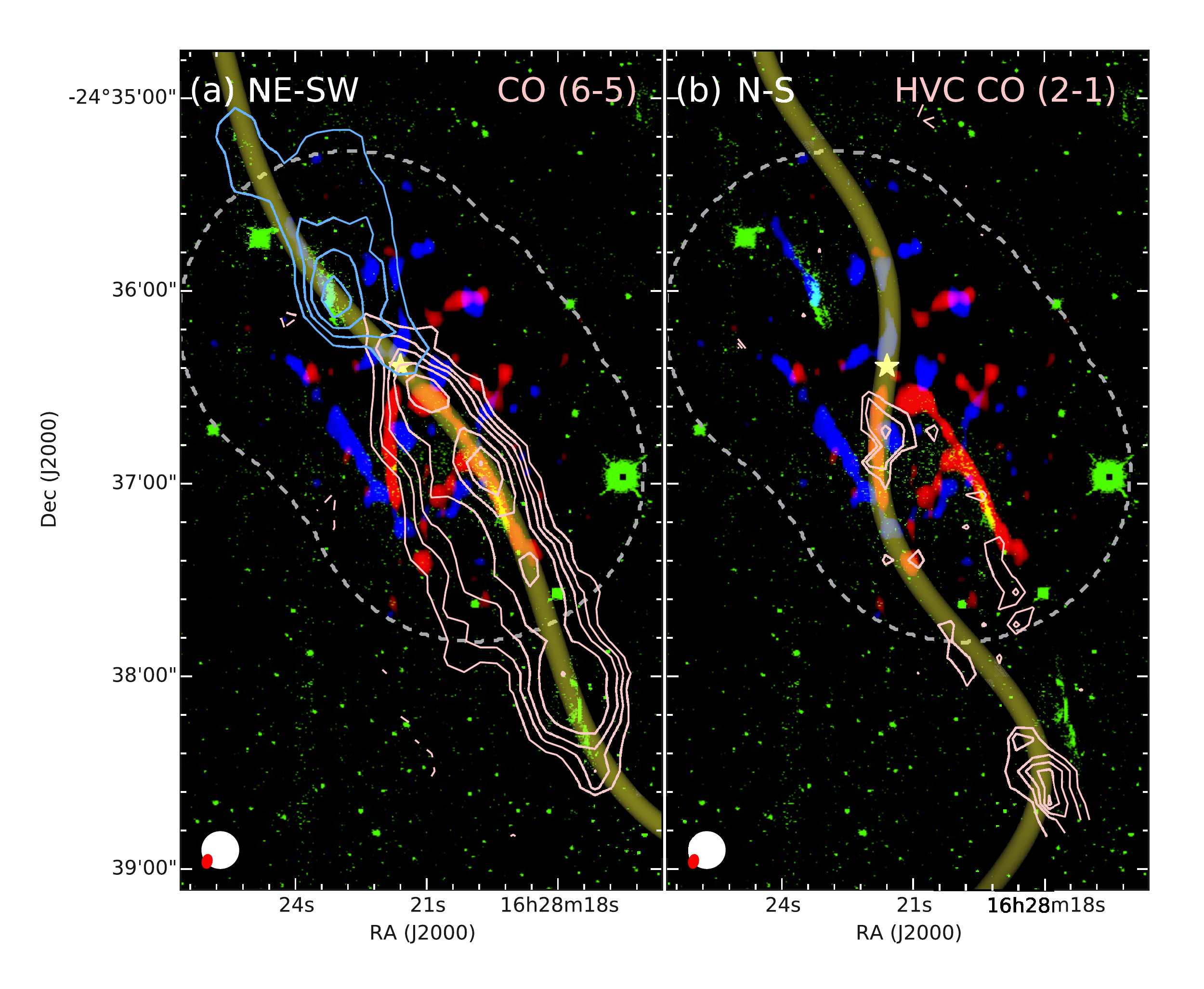}
\caption{ 
(a) Modeled precessing jet paths. The white star indicates the position of the infrared source and the yellow curves show the modeled jet paths considering only precession. 
The three-color images are the same maps as in Figure~\ref{fig:com} with an additional green color indicating the H$_2$ image.
The contours show the CO (6--5) maps as in Figure~\ref{fig:sin}.
(b) Same as the left figure but with the contours showing the high velocity IRAM CO (2--1) map (V $=$ 7.85 -- 9.05 km s$^{-1}$).
The contours start at 2$\sigma$ and increase by step of 1$\sigma$ with a rms noise level $\sigma=0.17$ K km s$^{-1}$.
}
\label{fig:locus}
\end{figure*}

\subsubsection{Precessing jets}
%Outflow locus fitting
\label{sec:pre}
The S-shaped jet driven by IRAS\,16253 was first detected by \citet{kh04} through H$_2$ emission at 2.12 $\mu$m (NE-SW).
S-shaped jets are believed to originate from precessing disks \citep{ma02,ra09,le10} which have tidal interactions with a close companion; in other words, IRAS\,16253 is likely a proto-binary system.
The SMA CO (2--1) and APEX CO (6--5) observations reveal the NE-SW jet locus and further connect the H$_2$ patches. 
Although the N-S ``jet'' is not associated with H$_2$ emission, it matches the component seen in the IRAM CO (2--1) map in a relatively high-velocity range of 7.85--9.05 km s$^{-1}$ (Figure \ref{fig:locus}b). In addition, it does not coincide with the edge of the cavity seen in scattered light (Figure~\ref{fig:smaco}).
Therefore, we suggest that the N-S component is a newly discovered jet driven by a binary component.
Furthermore, the IRAM CO (2--1) map covers a larger region than the SMA map and reveals a further extension and probably the head of the N-S jet (Figure~\ref{fig:locus}b). 
Given the absence of H$_2$ emission and absence/weakness of CO (6--5) emission in the N-S jet, the NE-SW and N-S jets must have very different physical conditions despite their similar environmental conditions in a common envelope.

We modeled the jet locus of both NE-SW and N-S jets by considering bipolar jets that originat from precessing disks.
An edge-on precessing jet can be described by a sinusoidal pattern with an amplitude increasing with the distance from the driving source \citep{ei96}.
Taking the outflow inclination into account, \citet{wu09} revised the formula from \citet{ei96} to analyse the precessing outflow driven by L1551\,IRS5 as
\begin{equation} 
\begin{bmatrix}
x\\
y
\end{bmatrix}
 = 
\begin{bmatrix}
\cos\psi & -\sin\psi \\
\sin\psi  & \cos\psi
\end{bmatrix}
\begin{bmatrix}
\alpha l \sin(2\pi l /\lambda_{\rm pre} + \phi_{\rm 0,\, pre})\\
l \cos(\theta_{\rm inc})\\
\end{bmatrix},
\label{eq:locus}
\end{equation}
%(which can be converted to half-opening angle of the precession cone, $\alpha'=\arctan(\alpha)$)
where $x$ and $y$ are the Cartesian coordinates, $\alpha$ is the precession amplitude, $\lambda_{\rm pre}$ is the precession spatial period, $l$ is the distance from the source, $\phi_{\rm 0,\, pre}$ is the initial phase at the source position, 
$\psi$ is the position angle of the jet symmetry (precession) axis in the plane of the sky, and $\theta_{\rm inc}$ is the inclination angle ($\theta=0$ for edge-on).
The distance $l$ in Equation (\ref{eq:locus}) is a true distance (as well as $\alpha$ and $\lambda_{\rm pre}$) rather than a projected distance due to $\cos(\theta_{\rm inc})$.
However, $\lambda_{\rm pre}$, $l$, and $\theta_{\rm inc}$ are degenerate
(only two free parameters), which prevents us from deriving the inclination angle.
Note that, although the inclination angle is included, Equation (\ref{eq:locus}) is only suitable for cases with small inclination angles; the sinusoidal pattern with an increasing amplitude is only an approximation of a helical pattern in side view.

Based on the outflow morphology,  we choose a nominal inclination angle of 20$\arcdeg$ (with respect to the plane of the sky), which is close to 15$\arcdeg$ estimated by \citet{ye15}, 
 in order to derive $\alpha$, $\lambda_{\rm pre}$, $\phi_{\rm 0,\, pre}$, and $\psi$. 
%The spatially extended and collimated outflow suggests that the jet/outflow most likely lies close to the plane of the sky.
%However, the outflow with almost no overlap between blue- and red-shifted emissions rejects an edge-on outflow. 
%\citep{ca90}.
The fact that the outflow is spatially extended and collimated implies that the jet/outflow most likely lies close to the plane of the sky. 
However, there is almost no overlap between the blue-shifted emission and the red-shifted emission, which excludes an edge-on configuration.
Our assumption of 20$\arcdeg$ is therefore a reasonable value.
The physical parameters affected by the choice of the inclination angle are discussed in Sections~\ref{sec:orbwig} and \ref{sec:outphy}.
The fitting process considers the H$_2$ emission, the APEX CO (6--5) outflow map, and the SMA CO (2--1) map for the NE-SW jet (Figure~\ref{fig:locus}a) and the IRAM 30 m high velocity map and SMA CO (2--1) map for the N-S jet (Figure~\ref{fig:locus}b).
We choose several peak positions along the jet to fit with Equation \ref{eq:locus}.
The best-fit loci of the NE-SW and N-S jets are shown in Figure~\ref{fig:locus} and the parameters of this best-fit model are listed in Table \ref{tab:prepa}.

\begin{figure}
\includegraphics[scale=.43]{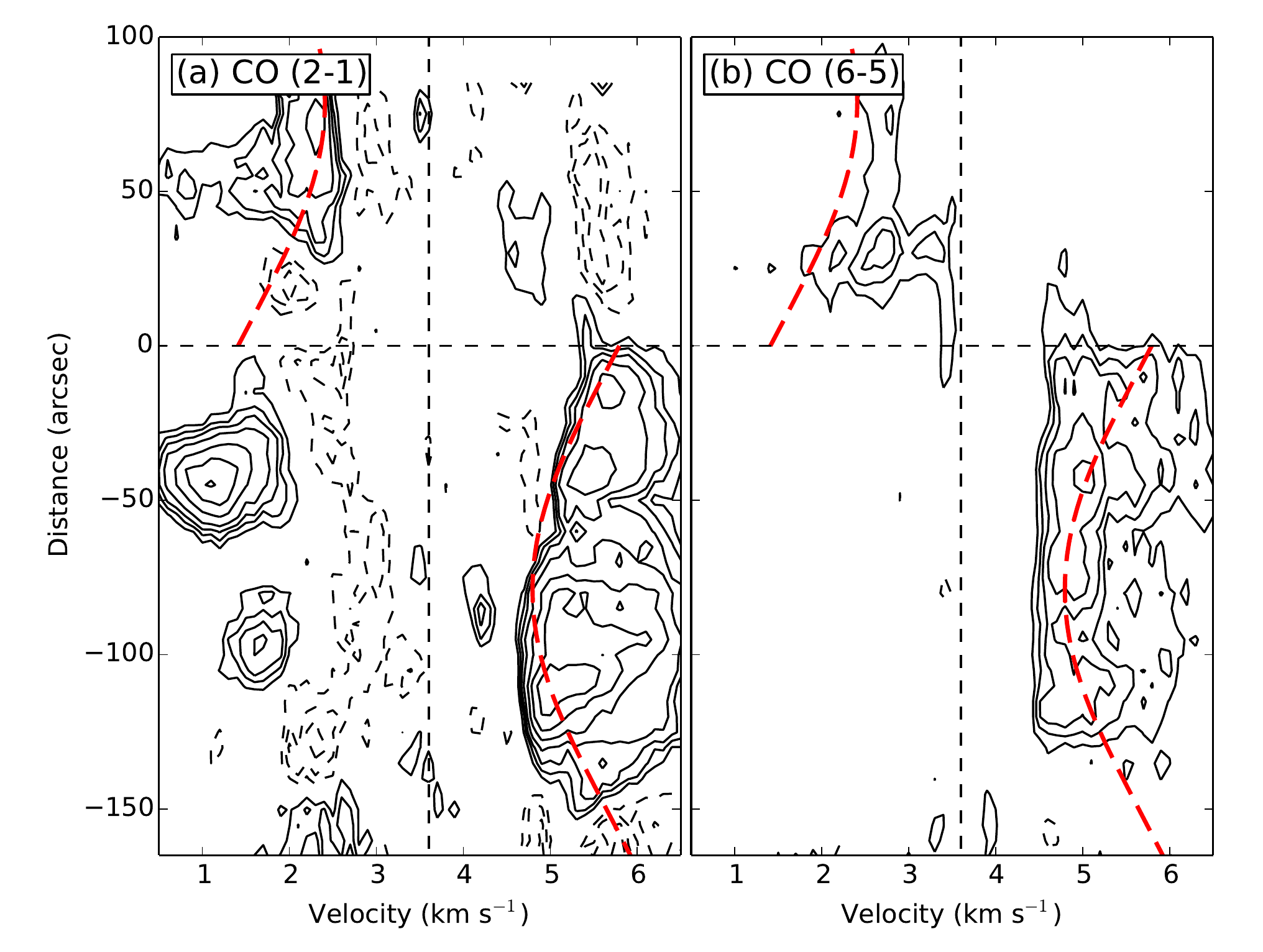}
\caption{Position-velocity diagram of the (a) IRAM CO (2--1) outflows and (b) APEX CO (6--5) outflows cut along the modeled loci shown in Figure~\ref{fig:locus}. The vertical axis shows the projected distance along a straight line with P.A. (East from North) of 38\fdg4 (the NE-SW jet), not along the (curved) modeled locus of the jet. 
The contour levels are -7, -5, -3, 3, 5, 7, 10, 15, 20, and 25$\sigma$ with $\sigma = 0.26$ K for CO (2--1) and $\sigma = 0.34$ K for CO (6--5). The black dashed lines indicate the source position and the systemic velocity. The red dashed lines show the best-fit precession model reported in Table \ref{tab:prepa}
}
\label{fig:pvsin}
\end{figure}

\begin{figure}
\includegraphics[scale=.7]{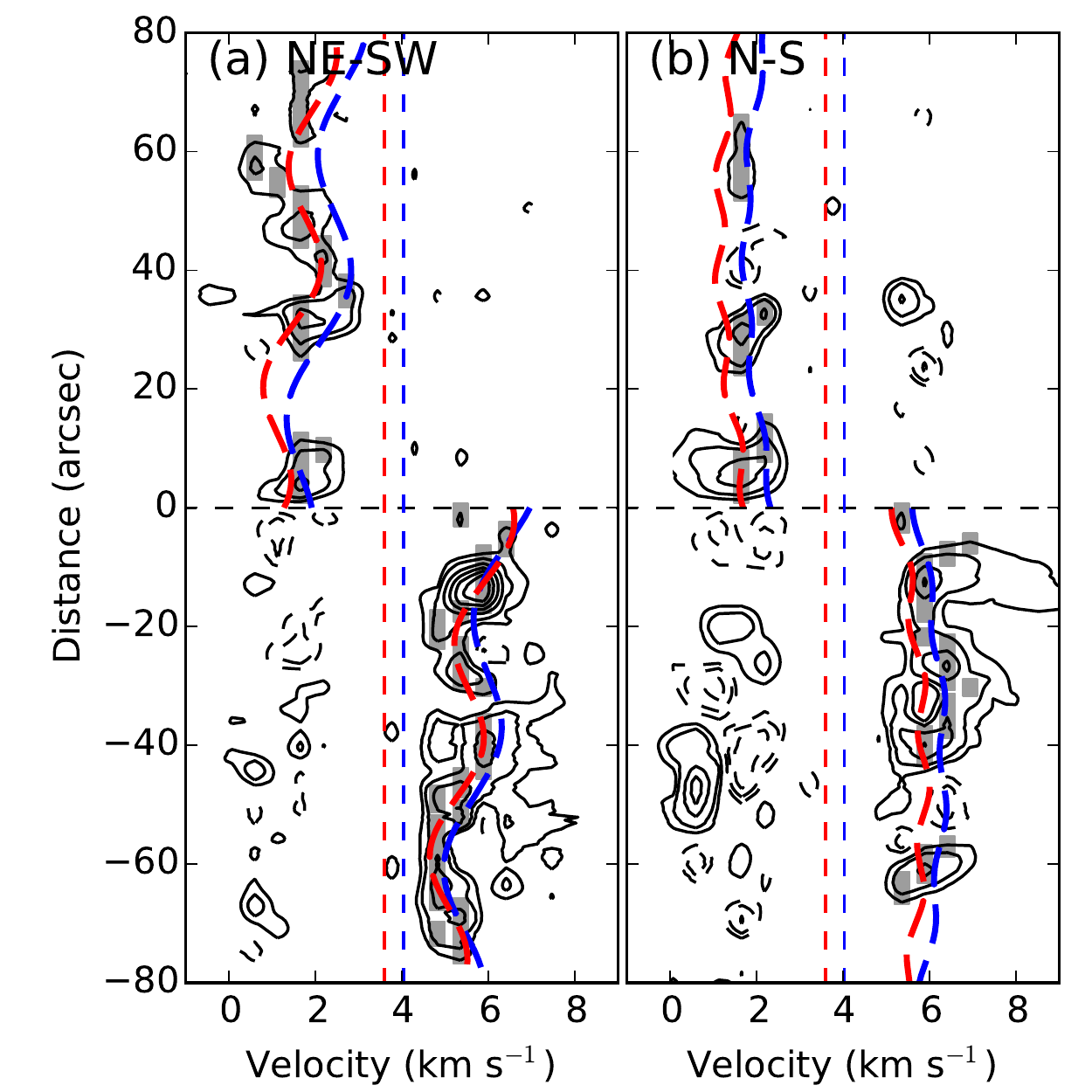}
\caption{SMA CO (2--1) position-velocity diagram of the (a) NE-SW and (b) N-S jets along the model paths shown in Figure~\ref{fig:locus}. The vertical axis shows the projected distance along P.A. $=$ 38.4$^\circ$ (a) and P.A. $=$ 0.7$^\circ$ (b). The contour levels are -20, -10, -5, 5, 10, 20, 30, 40 and 50$\sigma$ with $\sigma=0.06$ Jy beam$^{-1}$. 
The thick dashed lines show the best-fit models with a systemic velocity fixed to 4 km s$^{-1}$ (blue) 
and as a free parameter (red).  For the latter case, the best-fit systemic velocity is 3.5 km s$^{-1}$. In each case, the systemic velocity is marked with the thin dashed line of the same color. 
The grey bars indicate the position of the data points that were used for the fit.
}
\label{fig:pvsma}
\end{figure}

%PV-diagram for Single dish data
%With the locus in the position-velocity (PV) diagrams, we can further determine a jet velocity ($V_{\rm jet}$).
To trace the dynamics of the jets, we plot a special PV diagram with its PV cut along the jet locus 
from the precession models in Figure~\ref{fig:locus}.  
Note that the position in our PV diagram is the projected distance to the source along the jet symmetry axis in the plane of the sky (i.e., P.A. in Table \ref{tab:prepa}),
not the distance along the (curved) jet locus.
This enables us to convert the distance in the PV diagram into a dynamical age of the jet, which makes our following analysis easier especially for modeling the orbital motion of a binary system (Section \ref{sec:orbwig}).
%This procedure enables us to convert the distance to the dynamical time of the jet with the inclination angle and precession amplitude. % $\alpha.
Figure \ref{fig:pvsin} shows the PV diagram for the IRAM CO (2--1) and APEX CO (6--5) in the NE-SW jet.
% precessing jet model in PV diagram
The precessing jet in the PV diagram
can be described with an arc structure:
\begin{subequations} 
\begin{flalign}
{\rm D} = & l\cos(\theta_{\rm inc}), \label{eq: prevela}\\
V_{\rm LOS}  =  & V_{\rm LSR} \pm V_{\rm jet}[\cos(\alpha')\sin(\theta_{\rm inc}) \nonumber \\
& +\sin(\alpha')\cos(\theta_{\rm inc})\cos(2\pi l/\lambda_{\rm pre}+\phi_{\rm 0,\, pre})], \label{eq: prevelb}
\end{flalign}
\label{eq:prevel}%
\end{subequations}
where D is the projected distance to the source, $V_{\rm LOS}$ is the velocity along the line of sight, 
$\alpha'$ is the half-opening angle of the precession cone 
and can be obtained by $\alpha' = \arctan(\alpha)$ \citep{wu09}.
The positive and negative signs in Equation \ref{eq: prevelb} are used for the red- and blue-shifted lobes, respectively.
We have derived these parameters by fitting the jet locus (Equation \ref{eq:locus}) except for $V_{\rm jet}$ which is determined by the arc line position along the velocity axis in the PV diagram (Figure~\ref{fig:pvsin}).
We estimate a jet velocity from the CO (6--5) PV diagram since the APEX CO (6--5) map likely traces the NE-SW jet better than the IRAM CO (2--1) map.
However, the CO (6--5) map also contains the emission from the outflow cavities, which suggests that the jet velocity of 6.0 km s$^{-1}$ may be underestimated
Obtaining the jet velocity from the interferometer could be more reasonable because it is more sensitive to the smaller scales (see Section \ref{sec:orbwig}). 
We note that the distance in Equation \ref{eq: prevela} is an approximation which does not take into account the contribution of the precession motion that produces the velocity 
\begin{equation}
V_{\rm pre}=V_{\rm jet}\sin(\alpha')\cos(2\pi l/\lambda_{\rm pre}+\phi_{\rm 0,\, pre})
\label{eq:vpre}
\end{equation}
in Equation (\ref{eq: prevelb}).
The approximation is however reasonable for the small inclination angle $\theta_{\rm inc}$ and half-opening angle of the precession cone $\alpha'$.

\begin{deluxetable}{ccccccccc}
\tablewidth{0pt}
\tabletypesize{\tiny}
\tablecaption{Model Parameters of The Orbital Jets}
\tablehead{
\colhead{}    			& \colhead{$R$}	& \colhead{$\phi_{\rm 0,\, orb}$}	& \colhead{$V_{\rm jet}$}	& \colhead{$V_{\rm orb}$}	& \colhead{$V_{\rm LSR}$}\\
\colhead{Jets}      & \colhead{[arcsec]}	& \colhead{[deg]} & \colhead{km s$^{-1}$}	&
\colhead{km s$^{-1}$}	&	\colhead{km s$^{-1}$}
}
\startdata
\multicolumn{6}{c}{$\theta_{\rm inc}=20\arcdeg$}\\
\hline
NE-SW				& 0.45$\pm$0.02	& 326$\pm$4	& 7.3$\pm$0.05	& 0.52$\pm$0.02	&	3.5$\pm$0.01\\	
N-S					& 0.10$\pm$0.03  & 146$\pm$4	& 4.2$\pm$0.03	& 0.12$\pm$0.04	& 	3.5$\pm$0.01 \\
\hline
\multicolumn{6}{c}{$\theta_{\rm inc}=10\arcdeg$}\\
\hline
NE-SW				& 0.19$\pm$0.01	& 20$\pm$9	& 16.5$\pm$0.3	& 0.51$\pm$0.04	&  3.6$\pm$0.03 \\	
N-S					& 0.01$\pm$0.01 & 12$\pm$9	& 6.0$\pm$0.05	& 0.03$\pm$0.02	& 3.6$\pm$0.03\\
\hline
\multicolumn{6}{c}{$\theta_{\rm inc}=30\arcdeg$}\\
\hline
NE-SW				& 0.79$\pm$0.02	& 310$\pm$4	& 4.5$\pm$0.03	& 0.55$\pm$0.02	& 3.5$\pm$0.01\\	
N-S					& 0.02$\pm$0.05 & 130$\pm$4	& 3.4$\pm$0.03	& 0.02$\pm$0.04 & 3.5$\pm$0.01
\enddata
\tablecomments{Col.\ (1): Jet component. Col.\ (2): Radius of the binary orbital motion. Col.\ (3): Initial phase for orbital motion at the source. Col.\ (4): Jet velocity. Col.\ (6) Orbital Velocity of the corresponding driving source. 
Note that Col. (2) and (4) have a relation $\frac{R_1}{V_{\rm orb,\,1}}=\frac{R_2}{V_{\rm orb,\,2}}$ where the subscripts 1 and 2 denote the two jets. The two jets have a phase difference of $180\arcdeg$ due to the orbital motion of the binary system
}
\label{tab:orbpa}
\end{deluxetable}

\subsubsection{Orbital wiggling in the outflows}
\label{sec:orbwig}
%orbital model
%PV-diagram for interferometer data
In addition to the precession caused by the tidal interaction between the noncoplanar disk and the close companion,
the binary orbital motion can also affect the jet locus and produce a small-scale wiggling pattern
in both the jet map and PV diagram \citep{ra09, wu09, hi10}.
Figure~\ref{fig:pvsma} shows the PV diagram of the SMA CO (2--1) 
data for both the NE-SW and N-S jets along the jet locus 
in the same way as we did for the single-dish data (Figure~\ref{fig:pvsin}).
The SMA data have a higher angular resolution, but a lower spectral resolution and a smaller field of view than the single-dish data.

%% Start the new paragraph here.
We see in the SMA PV diagram small condensations that seem to oscillate in the velocity direction. We propose that these oscillations are due to the binary orbital motion.
Such oscillations are not seen in the PV diagrams of the single-dish data because the spatial resolution ($\sim$11\farcs2) is insufficient to resolve the wiggling pattern with a spatial period of $\sim$40\arcsec.
In order to study the wiggles in the SMA PV diagrams, we add the effect of orbital motion
\begin{equation}
V_{\rm orb,\, rad} = V_{\rm orb}\cos(\frac{V_{\rm orb}}{R}{\rm t}+\phi_{\rm0,\, orb})
\label{eq:vorb}
\end{equation}
into Equation (\ref{eq:prevel}) (also see Equation \ref{eq:vpre}):
\begin{equation}
V_{\rm LOS}  = V_{\rm LSR} \pm V_{\rm jet}\cos(\alpha')\sin(\theta_{\rm inc}) +(\pm V_{\rm pre}+V_{\rm orb,\, rad})\cos(\theta_{\rm inc}) \label{eq:vfint}%
\end{equation}
where $V_{\rm orb}$ is the orbital velocity, $V_{\rm orb,\, rad}$ is the projection of the orbital velocity 
along the line of sight if the inclination angle is 0,
R is the orbital radius, t is time, and $\phi_{\rm 0,\, orb}$ is the initial phase at the source position. 
To derive the orbital parameters, we fit the PV-diagram with Equation (\ref{eq:vfint}).
We select the peak velocity in each position as our data points in the PV-diagram, in which we chose a threshold of S/N$>$5.
We set the weighting for each data point as an integer calculated from w=(S/N)/5.
Finally, we artificially remove the points which are not associated to the jets.
The best-fit model and the data points are shown in Figure \ref{fig:pvsma} and its parameters are listed in Table \ref{tab:orbpa}.
Since the wiggling structure is better seen in the NE-SW jet than in the N-S jet,
the fitting process is mainly based on the NE-SW jet.
We first derive the jet velocity, orbital velocity, orbital radius, and initial phase of the NE-SW jet.
Consequently, only two free parameters (jet velocity and orbital velocity) remain undetermined in the N-S jet,
since the orbital velocity is proportional to the radius ($\frac{R_1}{V_{\rm orb,\,1}}=\frac{R_2}{V_{\rm orb,\,2}}$) and the phases have a difference of $180^\circ$ in a binary system.
Although the determinations of the data points is quite artificial in the fitting procedure, the significance of the orbital velocity ($V_{\rm orb}=0.53\pm0.02$ km s$^{-1}$, with more than 26$\sigma$ confidence level) for the NE-SW jet suggests that the wiggles produced by $V_{\rm orb}$ in Equation \ref{eq:vorb} and \ref{eq:vfint} are trustworthy.
For the N-S jet, the orbital velocity $0.1\pm0.03$ km s$^{-1}$ with the confidence level of $\sim$3$\sigma$ is insufficient to support the existence of oscillation.
Nevertheless, the orbital velocity and radius (as well as period) of the NE-SW jet-driving source is independent of the result of the fit of the N-S jet and should be reliable.
As a result, the best-fit model implies a very small binary separation of $\sim0\farcs55$ which is consistent with the SMA and JVLA continuum observations (see Section \ref{sec:cen})

Figure \ref{fig:pvsma} also includes the best-fit model assuming a systemic velocity of 4 km  s$^{-1}$ based on the N$_2$H$^+$ (1--0) observations toward the parent core \citep{to12,hs15}. 
This model is however not adopted since N$_2$H$^+$ (1--0) does not necessarily trace the systemic velocity of the central objects.
On the other hand, with the systemic velocity kept as a free parameter, the best-fit yields $V_{\rm LSR}\sim$3.5 km s$^{-1}$, which is close to the velocity of 3.4 km s$^{-1}$ found by \citet{st06} on the basis of the CO (3--2) outflow observations.

\begin{figure*}
\includegraphics[scale=.53]{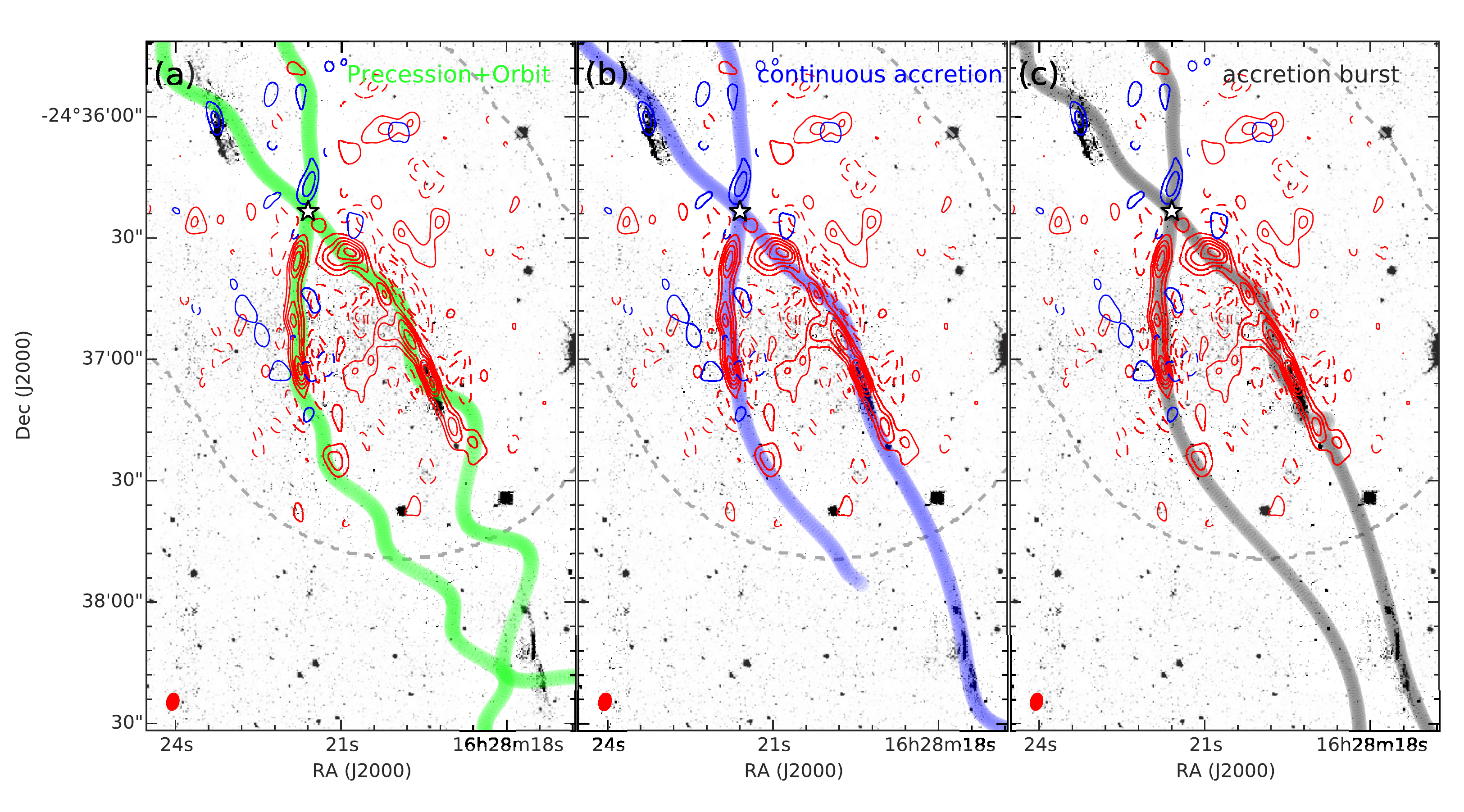}
\caption{Same as Figure~\ref{fig:locus}, but with models that also include the orbital motion.
The models assume that the mass of the central stars are constant (left), increase linearly (middle), and increase suddenly (right) with colors corresponding to the solid-line models in Figure~\ref{fig:pvmod}.
}
\label{fig:spamod}
\end{figure*}

\subsubsection{Orbital evolution and mass accretion models}
% for distant position, why we didn't see wiggling in spatial? Answer could be variation wiggling?
%variation of orbital velocity
\label{sec:locusmod}

The early evolution of binary systems has been a long-standing problem and is still poorly understood.
The evolution of the multiplicity frequency has recently been studied based on high resolution interferometric surveys \citep{re14}.
Several studies have resolved a number of multiple systems at the Class 0/I stage \citep{lo00,ch08,ch09,ma10,en11,to13}. The most thorough survey was presented in \citet{ch13}, and they found the multiplicity frequency of Class 0 protostars to be approximately twice higher than that of Class I sources and pre-main sequence stars, which could be interpreted as binary evolution (migration) in the early stages \citep{zh13}. Note that Chen et al.'s sample includes IRAS\,16253, but they did not resolve the binary system.

Our best-fit of the jet wiggling pattern in the PV-diagram (Figure~\ref{fig:pvsma}) leads to a problem: we do not see the wiggles due to the orbital motion in the jet in the spatial domain.
Figure \ref{fig:spamod}a shows the predicted spatial wiggling pattern corresponding to the best-fit model in the PV-diagram,
in which the southernmost patch of H$_2$ emission is not fitted.
We propose that a time-variable orbital motion could be the answer to this problem.
Here we provide several simple models to approach the problem considering migration and mass accretion.

\begin{figure*}
\includegraphics[scale=.87]{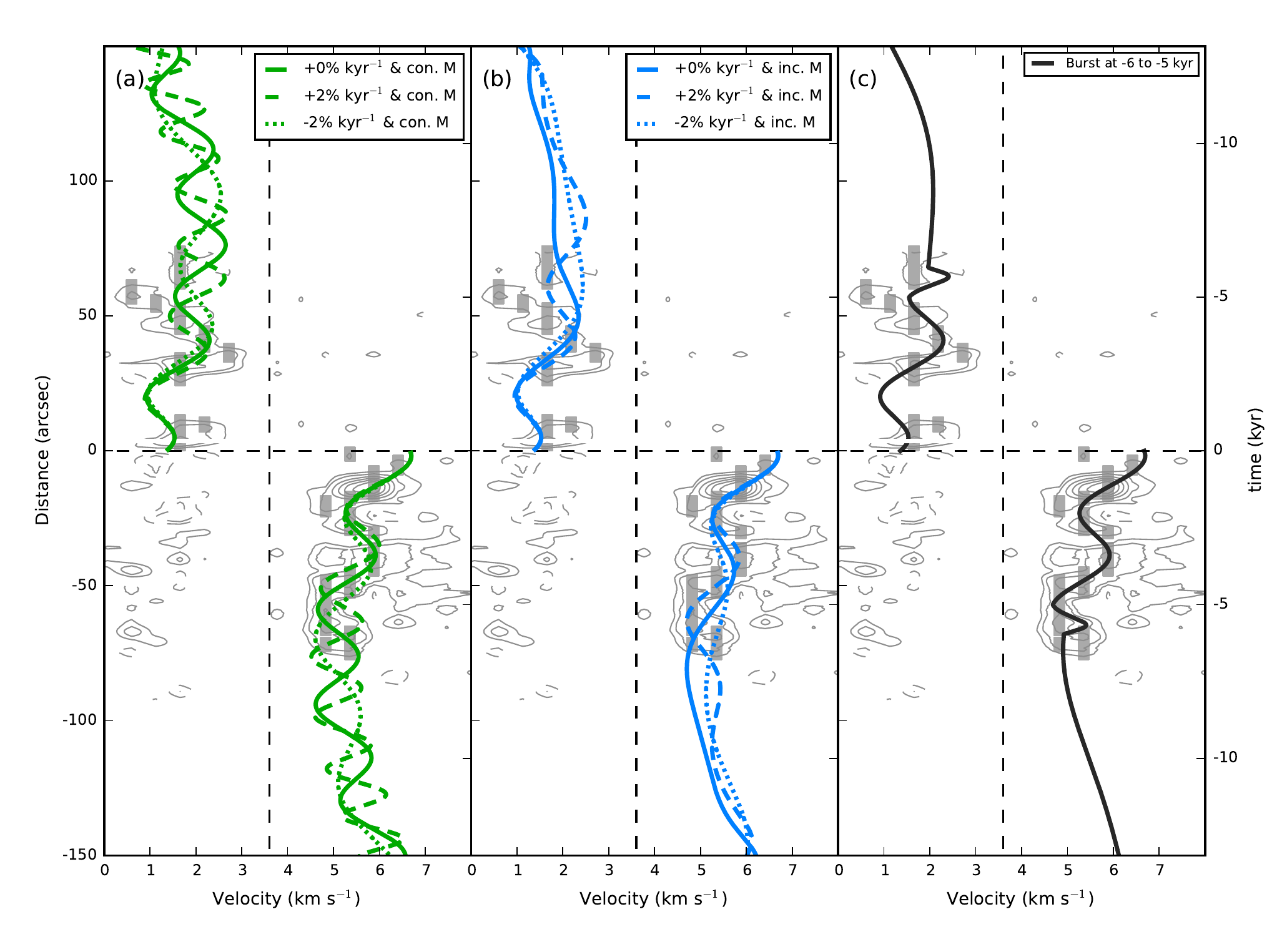}
\caption{Same as Figure~\ref{fig:pvsma} with additional time-dependent  models.
The models include separation increasing by 2\%, unchanged, and decreasing by 2\% per thousand year with constant mass (left) and increasing mass (middle).
For the increasing mass model, the total mass at t=$-$13 kyr was 10\% of the current mass,
and the mass increase rate is constant.
The right panel shows a model with an accretion burst between -6 kyr and -5 kyr
during which the mass increases linearly.
}
\label{fig:pvmod}
\end{figure*}

%model include vary binary separation and mass
We start from Kepler's 3rd law,
\begin{equation}
P_{\rm orb}^2 = \frac{4\pi^2a^3}{GM_{\rm tot}},
\label{eq:kep}
\end{equation}
where $P_{\rm orb}$ is the orbital period, $a$ is the binary separation, and $M_{\rm tot}$ is the total mass of the central objects.
Thus, the orbital velocity and period have the following relations with $M_{\rm tot}$ and $a$
\begin{equation}
V_{\rm orb} \propto (\frac{M_{\rm tot}}{a})^{1/2},~P_{\rm orb} \propto (\frac{a^3}{M_{\rm tot}})^{1/2}.
\label{eq:kepvar}
\end{equation}
Considering the binary separation $a$  (i.e.\ $R_1+R_2$) as a function of time $a({\rm t})$, we replace the orbital velocity and binary separation in Equation \ref{eq:vfint} by the time-dependent functions $V_{\rm orb}({\rm t})$ and $R({\rm t})$ assuming a constant total mass $M_{\rm tot}$.
Three models with time-dependent separation are shown in Figure \ref{fig:pvmod}a: one model with a constant orbital radius and two models with an orbital radius decreasing or increasing by 2\% per thousand year, respectively.
To study the evolution, Figure \ref{fig:pvmod} also shows the time axis on the right-hand side which is obtained assuming a jet velocity of 7.3 km s$^{-1}$ and an inclination angle of $20^\circ$ (see Table \ref{tab:orbpa}).
In comparison to the model with constant separation (solid line in Figure \ref{fig:pvmod}a), the model with an increasing separation clearly shows a growing period 
as the system evolves (in the direction toward the driving source position), while the model with a decreasing separation has an opposite behavior.

Because of the on-going accretion process onto the central objects, we actually expect $M_{\rm tot}$ to increase with time. Therefore, we replace the constant mass $M_{\rm tot}$ in Equation \ref{eq:kepvar} with a time-variable factor $M_{\rm tot}({\rm t})$. We set the models with an initial mass of 10\% of the current mass and a constant mass accretion rate over its dynamical time ($1.3\times10^4$ yr).
The results of these models are shown in blue in Figures~\ref{fig:spamod}b and \ref{fig:pvmod}b
in which a possible variation of the binary separation is also included (+2\%, 0\%, and -2\% per thousand year).
Furthermore, we present in Figures~\ref{fig:spamod}c and \ref{fig:pvmod}c another model to mimic an episodic accretion process with 90\% mass accreting between -6000 yr and -5000 yr (black lines).
We find that both continuous and episodic accretion models have the jet loci go through the southernmost H$_2$ patch (Figures~\ref{fig:spamod}b and \ref{fig:spamod}c).
However, all these modeled loci slightly deviate from the H$_2$ emission in the northern part of the NE-SW jet.
We suggest that the models with mass increments are possible approaches to the problem of absence of wiggling pattern but it may requires more complicated models or simulations than our simple description. 
Furthermore, it is likely that the modeled loci in the PV diagram (Figure \ref{fig:pvmod}) would allow us to determine the mass accretion history if high angular and spectral resolution data covering the whole jet are presented.
With such additional data, our models may provide an opportunity to study the mass accretion history and/or the orbital evolution (migration) in a binary system through the wiggling pattern of the molecular jets/outflows.

\subsection{Physical conditions}
\subsubsection{Central object}
\label{sec:cen}
% Central object propeties
%Continuum
The SMA continuum emission at 224 GHz toward the center is resolved with a deconvolved $FWHM$ size of $3\farcs4\times2\farcs3$. 
The continuum flux of the central object at 224 GHz $S_{\rm 1.3mm}$ is 37.7$\pm$0.15 mJy which is about 10\% less than the flux reported in \citet{ye15}.
If we adopt a dust opacity $\kappa_{\rm 1.3mm} = 0.01\,\rm cm^2\,g^{-1}$ \citep{os94} and assume a dust temperature $T_{\rm dust} = 15$ K, we obtain a dust mass of 0.018 $M_\odot$ around the central objects (0.032 $M_\odot$ if $T_{\rm dust} = 10$ K).
However, no elongated structure is found in our continuum map,
suggesting that the current resolution is not able to resolve the protostellar disk or a binary system.
Thus, it is still unclear whether the 224 GHz dust emission is associated to a protostellar disk, circumbinary disk, or the inner part of the protostellar 
envelope.

The JVLA continuum emission at 43 GHz is detected with a $FWHM$ size of $0\farcs8\times0\farcs6$ which is about the size of the beam.
The integrated flux density at 43 GHz is $S_{\rm 7mm}=0.158\pm0.01$ mJy.
If we use a graybody approximation $S_{\nu} \propto \nu^{2+\beta}$, we obtain an opacity index $\beta = 1.3$ from the SMA and JVLA data.
However, the angular resolutions and sensitivities of these two data sets are very different.
In addition, the extended emission at 224 GHz continuum is likely from more than a point source.
Thus, we cannot conclude if these two continuum emissions originate from the same source and what it is.

Though the continuum emissions at 43 GHz and 224 GHz do not show clear sign for a binary system,
the source sizes provide us clues about the binary separation.
The well detected dust continuum at 224 GHz suggests that the binary separation is at most $\sim$3\farcs3 (SMA beam).
The JVLA compact emission further implies either a binary separation less than $\sim$0\farcs7 
or a companion being too faint to be detected at 43 GHz in our current observations.
As a result, the binary separation is either about less than 0\farcs7 or between 0\farcs7 and 3\farcs3 with a faint companion at 43 GHz.
Note that the binary separation derived through fitting the wiggling pattern is $\sim0\farcs55$, which supports the first possibility.

%C18O %N2D+ Chemistry, 
Figure \ref{fig:sma} shows the integrated intensity maps of N$_2$H$^+$ (1--0) (from Tobin et al.\ 2012a), C$^{18}$O (2--1), and N$_2$D$^+$ (3--2).
N$_2$H$^+$ is depleted toward the center as seen in IRAM 04191 \citep{be04}. 
The depletion region is filled with C$^{18}$O emission, suggesting that the N$_2$H$^+$ depletion is caused by its destruction by CO \citep{ca12}.
The extended C$^{18}$O emission toward the source center has a deconvolved $FHWM$ size of $4\farcs2\times2\farcs8$ ($\sim$500\,au\,$\times$\,350\,au) and
a peak intensity of $\sim$1 Jy beam$^{-1}$.
This extent of C$^{18}$O emission most likely traces the inner region of the envelope.
The point-like N$_2$D$^+$ source is at a projected distance of $\sim$325\,au (2\farcs6) from the continuum source. This distance is not inconsistent with the constraints on the binary separation derived from the continuum emission above, but it is inconsistent with the binary separation derived from the analysis of the wiggle pattern in Section \ref{sec:orbwig}. Therefore the N$_2$D$^+$ source is unlikely to be the binary companion.
Furthermore, the high deuterium enhancement implies a gas temperature $\lesssim$ 20 K, which hints that the N$_2$D$^+$ condensation is unlikely to contain a powering source of the outflow.
Therefore, we suggest that the N$_2$D$^+$ source is a molecular condensation in the core or probably a third small starless component if it is associated with a continuum source.

%\begin{landscape}
\hspace{-3cm}
\tabletypesize{\scriptsize}
\tabcolsep=0.08cm
\begin{deluxetable*}{llccccccccc}
\tablewidth{0pt}
\tablecaption{Outflow parameters}
\tablehead{
\colhead{}    			& \colhead{}	& \multicolumn{3}{c}{SMA CO (2--1) NE-SW\tablenotemark{a}}	& \multicolumn{3}{c}{SMA CO (2--1) N-S\tablenotemark{a}}	& \multicolumn{3}{c}{IRAM$+$APEX\tablenotemark{b}}	%& \multicolumn{3}{c}{APEX CO (6--5)\tablenotemark{c}	}
\\
\cmidrule(lr){3-5} \cmidrule(lr){6-8} \cmidrule(lr){9-11} %\cmidrule(lr){12-14}
\colhead{}      &	\colhead{} & \colhead{Blue}	& \colhead{Red}	& \colhead{Total}	& \colhead{Blue}	& \colhead{Red}	& \colhead{Total}	& \colhead{Blue}	& \colhead{Red}	& \colhead{Total}	
\\
\colhead{Velocity range}      &	\colhead{km s$^{-1}$} & \colhead{0.3-3.0}	& \colhead{4.6-9.8}	& \colhead{Total}	& \colhead{-0.7-3.0}	& \colhead{4.6-10.9}	& \colhead{Total}	& \colhead{0.55-2.75}	& \colhead{4.45-7.35}	& \colhead{Total}	
}
\startdata
\multicolumn{11}{c}{$\theta_{\rm inc}=20\arcdeg$}\\
\tableline
Mass                                    & $\times$10$^{-4}$$M_\odot$
& $0.80_{-0.31}^{+1.04}$	& $2.95_{-1.16}^{+3.96}$	& $3.75_{-1.47}^{+4.99}$			%SMA NE-SW jet
& $0.83_{-0.33}^{+1.09}$	& $2.09_{-0.82}^{+2.73}$	& $2.92_{-1.14}^{+3.83}$		    %SMA N-S jet
& $1.78_{-0.00}^{+0.70}$	& $13.00_{-0.00}^{+1.40}$	& $14.80_{-0.00}^{+2.00}$	\\	    %IRAM+APEX jet
%& $0.41_{-0.06}^{+0.34}$	& $3.84_{-0.61}^{+3.19}$	& $4.25_{-0.68}^{+3.54}$		 \\   %APEX jet
Momentum                        & $\times$10$^{-3}$$M_\odot$ km s$^{-1}$
& $0.43_{-0.17}^{+0.56}$	& $1.96_{-0.77}^{+2.61}$	& $2.39_{-0.94}^{+3.18}$			%SMA NE-SW jet
& $0.51_{-0.20}^{+0.67}$	& $1.74_{-0.68}^{+2.28}$	& $2.25_{-0.88}^{+2.95}$		    %SMA N-S jet
& $0.71_{-0.00}^{+0.36}$	& $7.73_{-0.00}^{+1.06}$	& $8.45_{-0.00}^{+1.41}$		    \\%IRAM+APEX jet
%& $0.18_{-0.03}^{+0.15}$	& $1.19_{-0.19}^{+0.99}$	& $1.37_{-0.21}^{+1.15}$		 \\   %APEX jet
Outflow forces ($F_{\rm CO}$)           & $\times$10$^{-7}$$M_\odot$ km s$^{-1}$ yr$^{-1}$
& $0.81_{-0.32}^{+1.05}$	& $2.98_{-1.17}^{+3.98}$	& $3.79_{-1.49}^{+5.04}$			%SMA NE-SW jet
& $0.92_{-0.36}^{+1.19}$	& $1.79_{-0.70}^{+2.34}$	& $2.70_{-1.06}^{+3.54}$		    %SMA N-S jet
& $1.16_{-0.00}^{+0.59}$	& $5.93_{-0.00}^{+0.81}$	& $7.09_{-0.00}^{+1.39}$		    \\%IRAM+APEX jet
%& $0.31_{-0.05}^{+0.25}$	& $0.94_{-0.15}^{+0.78}$	& $1.25_{-0.20}^{+1.03}$		 \\   %APEX jet
Kinetic luminosity ($L_{\rm kin}$)              & $\times$10$^{-4}$$L_{\odot}$
& $0.80_{-0.31}^{+1.04}$	& $3.83_{-1.50}^{+5.08}$	& $4.63_{-1.82}^{+6.17}$			%SMA NE-SW jet
& $1.02_{-0.40}^{+1.32}$	& $2.80_{-1.10}^{+3.65}$	& $3.81_{-1.49}^{+4.98}$		    %SMA N-S jet
& $0.82_{-0.00}^{+0.51}$	& $6.12_{-0.00}^{+1.02}$	& $6.94_{-0.00}^{+1.53}$		   \\ %IRAM+APEX jet
%& $0.23_{-0.04}^{+0.19}$	& $0.56_{-0.09}^{+0.46}$	& $0.79_{-0.13}^{+0.65}$		 \\   %APEX jet
Mass loss rate   ($\dot{M}_{\rm loss}$)         & $\times$10$^{-8}$$M_\odot$ yr$^{-1}$
& $1.49_{-0.58}^{+1.93}$	& $4.50_{-1.77}^{+6.00}$	& $5.98_{-2.35}^{+7.92}$			%SMA NE-SW jet
& $1.49_{-0.58}^{+1.95}$	& $2.14_{-0.84}^{+2.80}$	& $3.63_{-1.42}^{+4.75}$		    %SMA N-S jet
& $2.90_{-0.00}^{+1.14}$	& $9.96_{-0.00}^{+1.04}$	& $12.90_{-0.00}^{+2.20}$		\\    %IRAM+APEX jet
%& $0.69_{-0.11}^{+0.57}$	& $3.03_{-0.49}^{+2.52}$	& $3.72_{-0.59}^{+3.09}$		 \\   %APEX jet
\tableline
\multicolumn{11}{c}{$\theta_{\rm inc}$ dependent}\\
\tableline
Mass                                    & $\times$10$^{-4}M_\odot$
& $0.80_{-0.31}^{+1.04}$	& $2.95_{-1.16}^{+3.96}$	& $3.75_{-1.47}^{+4.99}$			%SMA NE-SW jet
& $0.83_{-0.33}^{+1.09}$	& $2.09_{-0.82}^{+2.73}$	& $2.92_{-1.14}^{+3.83}$		    %SMA N-S jet
& $1.78_{-0.00}^{+0.70}$	& $13.00_{-0.00}^{+1.40}$	& $14.80_{-0.00}^{+2.00}$	\\	    %IRAM+APEX jet
%& $0.41_{-0.06}^{+0.34}$	& $3.84_{-0.61}^{+3.19}$	& $4.25_{-0.68}^{+3.54}$		 \\   %APEX jet
Momentum                        & $\times$10$^{-3}\frac{1}{\sin(\theta_{\rm inc})}M_\odot$ km s$^{-1}$
& $0.15_{-0.06}^{+0.19}$	& $0.67_{-0.26}^{+0.89}$	& $0.82_{-0.32}^{+1.09}$			%SMA NE-SW jet
& $0.17_{-0.07}^{+0.23}$	& $0.60_{-0.23}^{+0.78}$	& $0.77_{-0.30}^{+1.01}$		    %SMA N-S jet
& $0.24_{-0.00}^{+0.12}$	& $2.65_{-0.00}^{+0.35}$	& $2.89_{-0.00}^{+0.48}$		\\    %IRAM+APEX jet
%& $0.06_{-0.01}^{+0.05}$	& $0.41_{-0.06}^{+0.34}$	& $0.47_{-0.07}^{+0.39}$		 \\   %APEX jet
Outflow forces ($F_{\rm CO}$)           & $\times$10$^{-7}\frac{\cos(\theta_{\rm inc})}{\sin^{2}(\theta_{\rm inc})}M_\odot$ km s$^{-1}$ yr$^{-1}$
& $0.10_{-0.04}^{+0.13}$	& $0.37_{-0.15}^{+0.50}$	& $0.47_{-0.18}^{+0.63}$			%SMA NE-SW jet
& $0.11_{-0.04}^{+0.15}$	& $0.22_{-0.09}^{+0.29}$	& $0.34_{-0.13}^{+0.44}$		    %SMA N-S jet
& $0.15_{-0.00}^{+0.07}$	& $0.74_{-0.00}^{+0.10}$	& $0.88_{-0.00}^{+0.18}$		 \\   %IRAM+APEX jet
%& $0.04_{-0.01}^{+0.03}$	& $0.12_{-0.02}^{+0.10}$	& $0.15_{-0.02}^{+0.13}$		 \\   %APEX jet
Kinetic luminosity ($L_{\rm kin}$)              & $\times$10$^{-4}\frac{\cos(\theta_{\rm inc})}{\sin^{3}(\theta_{\rm inc})}L_{\odot}$
& $0.03_{-0.01}^{+0.04}$	& $0.16_{-0.06}^{+0.22}$	& $0.20_{-0.08}^{+0.26}$			%SMA NE-SW jet
& $0.04_{-0.02}^{+0.06}$	& $0.12_{-0.05}^{+0.16}$	& $0.16_{-0.06}^{+0.21}$		    %SMA N-S jet
& $0.03_{-0.00}^{+0.02}$	& $0.26_{-0.00}^{+0.04}$	& $0.29_{-0.00}^{+0.07}$		 \\   %IRAM+APEX jet
%& $0.01_{-0.00}^{+0.01}$	& $0.02_{-0.00}^{+0.02}$	& $0.03_{-0.01}^{+0.03}$		 \\   %APEX jet
Mass loss rate   ($\dot{M}_{\rm loss}$)         & $\times$10$^{-8}\frac{\cos(\theta_{\rm inc})}{\sin(\theta_{\rm inc})}M_\odot$ yr$^{-1}$
& $0.54_{-0.21}^{+0.71}$	& $1.64_{-0.65}^{+2.19}$	& $2.18_{-0.86}^{+2.89}$			%SMA NE-SW jet
& $0.54_{-0.21}^{+0.71}$	& $0.78_{-0.31}^{+1.02}$	& $1.32_{-0.52}^{+1.73}$		    %SMA N-S jet
& $1.06_{-0.00}^{+0.41}$	& $3.62_{-0.00}^{+0.39}$	& $4.68_{-0.00}^{+0.80}$		 %\\   %IRAM+APEX jet
%& $0.25_{-0.04}^{+0.21}$	& $1.10_{-0.17}^{+0.92}$	& $1.35_{-0.21}^{+1.13}$		 \\   %APEX jet
\enddata
\tablecomments{
}
\tablenotetext{a}{The outflow parameters are derived by assuming an H$_2$ density of 10$^5$ cm$^{-3}$ and a $T_{\rm kin}$ of 40 K. The upper and lower limits correspond to calculations assuming a lower $T_{\rm kin}$ of 30 K and a higher $T_{\rm kin}$ of 50 K.}
\tablenotetext{b}{The outflow parameters are derived with CO (2--1)/(6--5)/(7--6) intensities from the region where the CO (2--1) intensity has a signal-to-noise ratio above 3. While the mass is derived by assuming an H$_2$ density of 10$^5$ cm$^{-3}$, the uncertainties are obtained from the masses derived with $n_{\rm H_2} = 10^3$ cm$^{-3}$ (upper limit) and $n_{\rm H_2} = 10^7$ cm$^{-3}$ (lower limit). The pixels with derived kinetic temperature lower than 12 K are removed from the outflow mass estimations.}
%\tablenotetext{c}{The outflow parameters are derived with CO (6--5) intensities from where CO (6--5) intensities with S/N $>$ 3 and assuming a $T_{\rm kin}$ of 120 K.The uncertainties are obtained from the derived masses by assuming $T_{\rm kin} = 70$ K (upper limit) and $T_{\rm kin} = 170$ K (lower limit).}
\label{tab:outpa}
\end{deluxetable*}

%\end{landscape}

\subsubsection{Outflow physical parameters}
\label{sec:outphy}
% Calculation with RADEX
We derive the gas temperature ($T_{\rm gas}$) and column density ($N_{\rm CO}$) toward the large-scale CO outflow with the single-dish CO (2--1)/(6--5)/(7--6) maps.
We smoothed these maps to the same angular resolution (11\farcs2) and constructed data cubes with the same spatial and spectral grids (see Sections \ref{sec:iramobs} and \ref{sec:apexobs}).
We thus obtain the CO (2--1)/(6--5)/(7--6) intensities at each given position and velocity.
Although CO (7--6) is undetected in most regions, the upper limit could still provide constraints on the physical properties.
We use the non-LTE radiative transfer code RADEX \citep{va07} to derive the CO column density and gas temperature in each cell ($5''\times5''\times0.1\,{\rm km\,s^{-1}}$). 
In addition to the column density and gas temperature, there are two free parameters: the H$_2$ density ($n_{\rm H_2}$) and the CO line width.
We assume a H$_2$ density of $10^{5}$ cm$^{-3}$ which is about the critical density ($n_{\rm cri}\sim1.2\times10^{5}$ cm$^{-3}$) of CO (6--5) \citep{ya10,yi13} and we later use $n_{\rm H_2}=10^3$ cm$^{-3}$ and $n_{\rm H_2}=10^7$ cm$^{-3}$ to derive the uncertainties (upper and lower limits).
We set the line width (dispersion) as $\sigma=\frac{\Delta V_{\rm chan}}{\sqrt{2\pi}}$ such that the ``integrated'' opacity $\int\tau_v dv=\tau_{\rm peak}\sqrt{2\pi}\sigma$ is equal to $\tau_{\rm peak} \Delta V_{\rm chan}$ where $\Delta V_{\rm chan}$ is the channel width of 0.1 km\,s$^{-1}$.
This setup enables us to estimate the column density and gas temperature in each channel using the output radiation temperatures of RADEX.

% Result and two type of maps (with 2-1/6-5/7-6 and only 6-5)
With the aforementioned method, we derive the column density and gas temperature in each pixel in the channel maps.
We adopt a signal-to-noise ratio threshold of 3 in CO (2--1)
to select the region where the outflow is detected.
As a result, we obtain  ``channel column density'' and ``channel gas temperature'' maps. 
The column densities are further used to calculate the outflow mass, momentum, force ($F_{\rm CO}$), kinetic luminosity ($L_{\rm kin}$), and mass-loss rate ($\dot{M}$). The results are given in Table \ref{tab:outpa}.
The upper and lower limits of column density are calculated with the H$_2$ density assumptions of $10^{3}$ cm$^{-3}$ and $10^{7}$ cm$^{-3}$, respectively.
The lower limits derived with $n_{\rm H_2}=10^7$ cm$^{-3}$ (i.e. under LTE conditions) are very close to the values obtained with 
$n_{\rm H_2}=10^5$ cm$^{-3}$ which is close to the critical density of CO (6--5).
For the upper limits derived under non-LTE conditions ($n_{\rm H_2} \ll n_{\rm cri}$), higher column densities are required to reproduce the same observed brightness temperature.

\subsubsection{Warm gas properties}
Mid-J CO  transitions are good tracers of warm gas in molecular outflows compared to low-J transitions \citep{go13}.
Since we have found that CO (2--1) and CO (6--5)/(7--6) probe different gas (Section \ref{sec:sin}) and
the step 3 summation maps likely trace the collimated jet better (Section \ref{sec:sca}), we use the CO (6--5) and (7--6) maps and their step 3 summation maps to study the shocked gas.
Figures \ref{fig:rat7665}a and \ref{fig:rat7665}b show the maps of the integrated intensity ratios of CO (7--6) to CO (6--5) and that from the step 3 summation maps, respectively.
Excluding the region around the driving sources, the integrated intensity ratios are in the ranges of $\sim$0.3--0.7 in Figure \ref{fig:rat7665}(a) and $\sim$0.5--0.9 in Figure \ref{fig:rat7665}(b) from where CO (7--6) is detected.
This ratio seems to be relatively high in the H$_2$ emission regions.
Figure \ref{fig:radex} shows the modeled CO (7--6)/(6--5) intensity ratio as a function of H$_2$ density and kinetic temperature from RADEX \citep{va07} at an assumed CO column density of $10^{14}$cm$^{-2}$.
In the regions with $n_{\rm H_2}\lesssim10^{6}$ cm$^{-3}$, the intensity ratio is highly dependent on not only the gas temperature but also the H$_2$ density.
This implies a model degeneracy because we aim to derive three parameters (T$_{\rm gas}$, $n_{\rm H_2}$, and N$_{\rm CO}$) with only two data points (CO 6--5 and 7--6).
To break the degeneracies, we take the canonical CO abundance of $8\times10^{-5}$ and assume a uniform density distribution. We obtain the following relation:
\begin{equation}
(\frac{N({\rm CO})}{10^{16} \rm cm^{-2}})=0.12(\frac{n_{\rm H_2}}{10^{3} \rm cm^{-3}}) (\frac{d_{\rm LOS}}{\rm1000 au}),
\label{eq:geo}
\end{equation}
where $d_{\rm LOS}$ is the depth along the line of sight. 
If we take the jet knot size of a few thousands au as $d_{\rm LOS}$ as seen in Figure \ref{fig:sinsca}, the warm gas associated with the jet knot has a low H$_2$ density of $10^3-10^4$ cm$^{-3}$.
This result hints at a temperature of few hundreds Kelvin in the jet knot, which is lower than the hot gas temperature T$\sim$1000\,K derived from the infrared rotational transitions of H$_2$ by \citet{ba10}.

\section{Discussion}
\subsection{Physical properties of the outflows}
\subsubsection{Origin of warm gas: CO (6--5)/(7--6)}
\label{sec:ori}
%The origin of warm gas is highly debated van Kempen 2009a
The origin of warm gas and its heating mechanisms are still debated in low-mass protostars \citep{ke09b,di09}.
The possible options are summarized by \citet{di09} as: 
(1) heating of the collapsing core by the protostellar luminosity within $\sim$100 au;
(2) active heating in the shock created by the interaction of the jet/outflow with the envelope; 
(3) heating by UV photons along the outflow cavity wall \citep{sp95}; 
(4) heating of a forming protoplanetary disk by accretion shocks \citep{ce00}.

For the large-scale CO (6--5) outflow, options (1) and (4) are excluded, 
since they predict a spatially unresolved emission toward the source center.
We have decomposed the CO (6--5) outflow into two components (Section \ref{sec:sca}): (1) the large-scale structure matching the outflow cavity probed by IRAC 1 and (2) the small-scale structure that lies along the H$_2$ jet emission (Figure \ref{fig:sinsca}).
The large-scale structure is likely due to heating by UV photons.
However, the photon dominated region created by the central accretion disk is expected to have a size up to few thousands au \citep{sp95,ke09a}, which is obviously too small to explain the extent of the large-scale CO (6--5) emission ($\sim$150\arcsec, i.e. $\sim$19000 au). 
\citet{ke09a} suggested that the UV photons can also be produced in the bow shock regions in the outflow if $J$-shocks are present \citep{ne89}.
This interpretation is supported by the existence of $J$-shocks suggested by the H$_2$ line study \citep{ba10}.

\begin{figure}
\includegraphics[scale=.4]{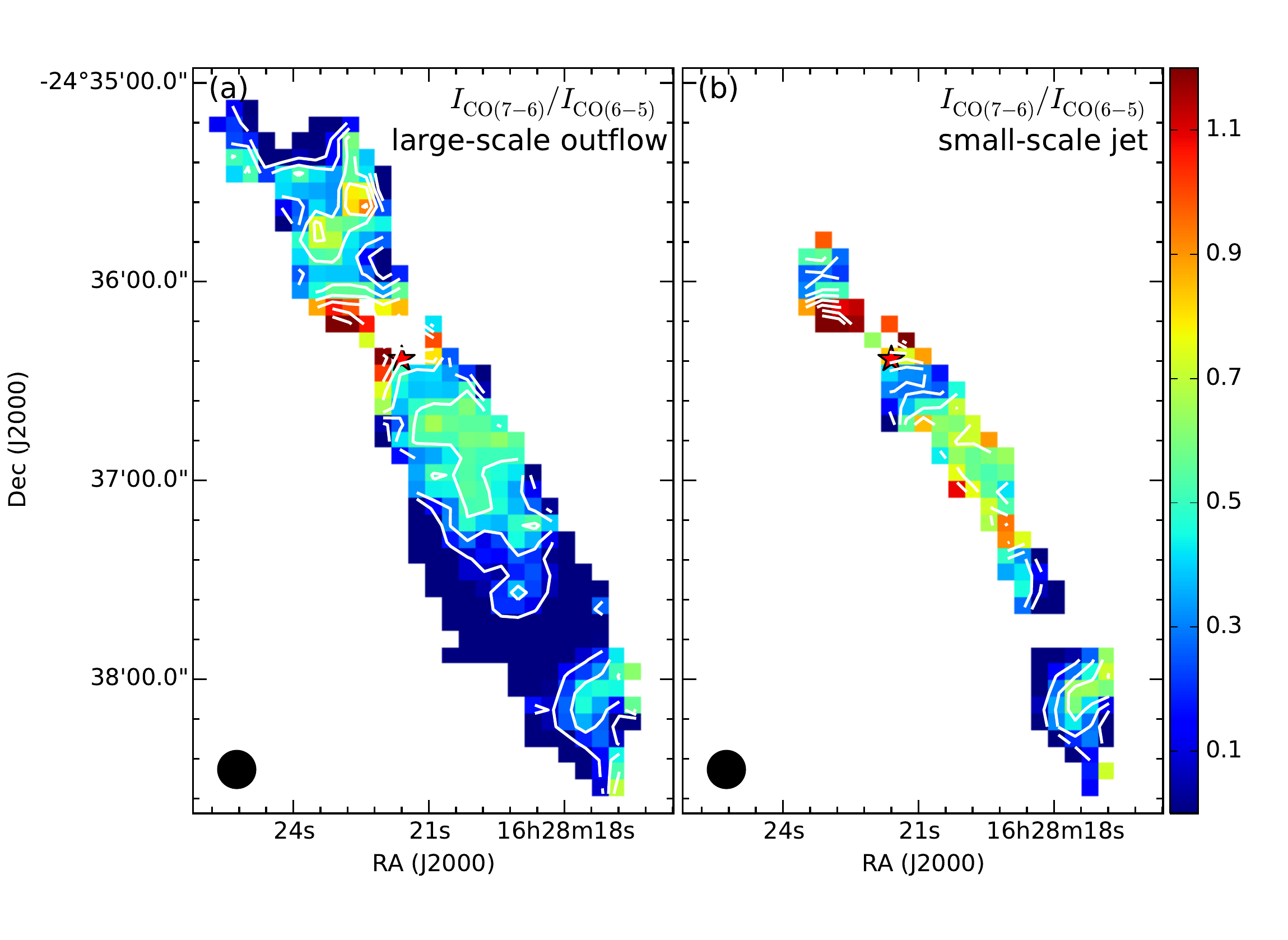}
\caption{(a) Map of the integrated intensity ratio CO (7--6)/(6--5) computed from the maps shown in Figure~\ref{fig:sin}. The ratio was computed only for the positions with a CO (6--5) integrated intensity with a signal-to-noise ratio above 3.
The contour levels are 0.1, 0.3, 0.5, 0.7, 0.9, and 1.1. (b) Same as (a) but with the step 3 summation maps (more sensitive to the jet components, see solid contours in Figure~\ref{fig:sinsca}) for both CO (6--5) and CO (7--6).}
\label{fig:rat7665}
\end{figure}

\begin{figure}
\includegraphics[scale=.55]{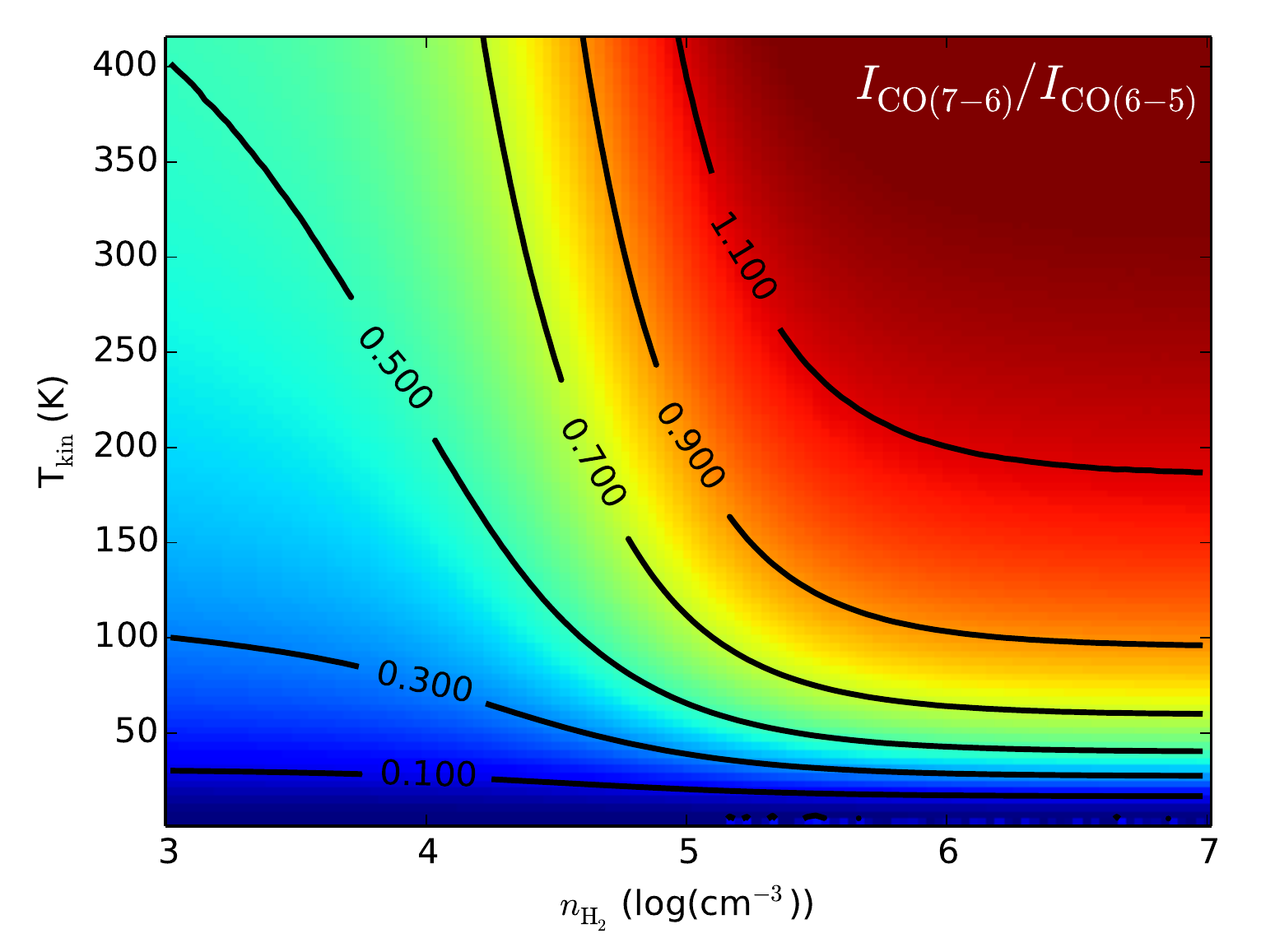}
\caption{Synthetic CO (7--6)/(6--5) line intensity ratio as a function of gas temperature and H$_2$ density obtained with RADEX \citep{va07}.
The CO column density is assumed to be $10^{14}~{\rm cm}^{-2}$.
}
\label{fig:radex}
\end{figure}

Although the large-scale CO (6--5) outflow emission mainly originates from UV heating, the knot-like structure matching the H$_2$ jet (Figure \ref{fig:sinsca}) is likely contributed by active heating in the shock regions.
By multiresolution decomposition, we extracted the small-scale component from CO (6--5) emission (in step 3) and it matches well the H$_2$ jet.
We conclude from this that these small-scale components may be associated to the jet shocks but high-angular-resolution observations are required for confirmation.
In addition, these CO (6--5)/(7--6) jet knots preferentially appear after the H$_2$ emission tips (Figure \ref{fig:sinsca}).
We speculate that CO is dissociated at the bow shock tip where H$_2$ is excited.
%{\color{red}[remove a sentence]}

The unresolved low velocity component in CO (7--6) (3.3--4.1 km s$^{-1}$, see Figure \ref{fig:APEXch1b}) toward the source center can be interpreted by either option (1) or (4).
Our current observations only allow us to constrain an upper limit  to the size of the CO (7--6) compact object as $<7.8\arcsec$ ($\sim$1000 au, the beam size without smoothing).
This prevents us from determining the origin of this component.
We note that this unresolved low velocity component is surprisingly not detected in CO (6--5). This casts some doubts about the existence of this component. Observations with higher angular resolution and sensitivity are required to solve this issue.

\subsubsection{Physical conditions in the outflows}
%introduction
We derived the outflow parameters in IRAS\,16253 using RADEX (Table \ref{tab:outpa}).
These parameters are derived from two data sets with different assumptions: (1) SMA CO (2--1) NE-SW and N-S jets with $n_{\rm H_2}=10^5$ cm$^{-3}$ and $T_{\rm kin}=40$ K, and (2) IRAM CO (2--1) and APEX CO (6--5)/(7--6) maps (hereafter IRAM$+$APEX) with $n_{\rm H_2}=10^5$ cm$^{-3}$.
The assumption of density $n_{\rm H_2}=10^5$ cm$^{-3}$ implies physical conditions close to LTE.
Based on the analysis of IRAM$+$APEX, we found an outflow temperature of around 40 K and we used it to derive the outflow parameters with the SMA CO (2--1) observation.

%outflow mass....
From IRAM$+$APEX, we estimate a total outflow mass of $14.8^{+2.0}_{-0}\times10^{-4} M_\odot$ 
(the upper and lower limits are derived from  $n_{\rm H_2}=10^3$ and $10^7$ cm$^{-3}$) which is comparable to that in \citet{st06}.
Assuming a gas temperature of 30 K,  \citet{st06} derived an outflow mass of $9.6\times10^{-4} M_\odot$ from the JCMT CO (3--2) observations in the optically thin limit;
since their CO (3-2) observations suffer from optical depth effects, \citet{st06} scaled the mass to $33.7\times10^{-4} M_\odot$ by assuming an optical depth correction of $\tau_{\rm CO (3-2)}/(1-e^{-\tau_{\rm CO (3-2)}})=3.5$.
However, this correction of optical depth could be quite uncertain and/or very different from source to source.
Our estimate based on multi-transition observations is likely more accurate.
Furthermore, \citet{ma13} used seven different methods to calculate the outflow forces of low-mass protostars in Ophiuchus, including IRAS\,16253, based on the JCMT CO (3--2) maps and our derived force is at about the median of the seven values.

%outflow parameters force, kinetic luminosity
We here compare the outflow parameters derived from SMA and IRAM$+$APEX.
Since the mapping areas are different between SMA and IRAM$+$APEX observations, the outflow mass and momentum cannot be directly compared.
Thus, we focus on the outflow force, kinetic luminosity, and mass loss rate which should be independent of the mapping areas. 
The SMA outflow forces and kinetic luminosities are comparable with that from IRAM$+$APEX,
if we sum up the contributions from both NE-SW and N-S jets (see Table \ref{tab:outpa}).
%, the SMA outflow forces and kinetic luminosities are consistent with that of IRAM$+$APEX.
We assume that the collimated jets seen with the SMA are driven by the central protostars and the large-scale outflows seen with IRAM$+$APEX are mostly from the entrained gas and/or outflow cavity wall. Based on the results given in Table~\ref{tab:outpa}, we then suggest that the collimated jets provide sufficient energy to drive the large-scale outflows.

\subsection{Proto-brown dwarf binary candidate}
% brown dwarf implication
Identifying a proto-BD is very difficult but important for understanding the mechanisms of BD formation (see Section \ref{sec:int}).
Here we discuss the parent core mass (Section \ref{sec:cor}) and the current stellar mass (Section \ref{sec:cendis}), and then conclude that IRAS-16253 may have insufficient mass to form a hydrogen-burning star ($>0.075M_{\odot}$). We further use the outflow force to support this and compare with other currently known young protostars and proto-BD candidates (Section \ref{sec:out}, see also Section \ref{sec:outMacc}). 
%We finally discuss the brown-dwarf forming mechanisms based on our results (section \ref{sec:eje}).

\subsubsection{Parent core mass}
\label{sec:cor}
%parent core mass
The growth of an accreting protostar depends on the mass in the parent core.
The core mass of IRAS\,16253 was estimated in previous studies (0.15 M$_\odot$ by Barsony et al.\ 2010; 0.2 M$_\odot$ by Stanke et al.\ 2006; 0.5 M$_\odot$ by Enoch et al.\ 2008; 0.8 M$_\odot$ by Tobin et al.\ 2012).
The inconsistencies between these results are due to different assumptions of temperatures and core sizes.
Although the mass of 0.8 M$_\odot$ derived by \citet{to12} using 8 $\mu$m extinction map is temperature independent, their core size with a diameter of 0.1 pc ($\sim$165\arcsec) is likely overestimated; this size is much larger than both the size derived by the COMPLETE project \citep{ri06} from the 850 $\mu$m map (Figure \ref{fig:smaco}b) and the $FWHM$ size of 53\arcsec$\times$43\arcsec~estimated by \citet{st06} from their 1.2~mm map.
Since the source is embedded in the $\rho$ Ophiuchus molecular cloud, the mass estimation considering a larger area can include a lot of material from the ambient cloud that may not participate in the accretion process.

One of the uncertainty in the core mass comes from the assumed dust temperature; \citet{en09} derive a mass of 0.51 M$_\odot$ at 1.1 mm with $T_{\rm dust}=15$ K and \citet{st06} estimate a mass of 0.2 M$_\odot$ with $T_{\rm dust}=20$~K at 1.2 mm.
We derive a dust temperature of $17.4 \pm 3.8$ K using the fluxes at 70, 100, 160, 350, 1100, 1200~$\mu$m from \citet{du08} by fitting a graybody model.
This temperature is consistent with the values of both \citet{en09} and \citet{st06} within the uncertainty. 
Rescaling the masses in \citet{en09} and \citet{st06} with a dust temperature of 17.4 K yields 0.41 M$_\odot$ and 0.24 M$_\odot$, respectively.
Besides, the dust opacity used for the core mass estimate could be very uncertain.
\citet{en09} adopted the widely used theoretical opacity of \citet{os94} which has an deviation may not more than a factor of 2.}
We therefore consider 0.82 M$_\odot$ as an upper limit of the core mass.
If we assume a star formation efficiency (SFE) of 0.1--0.3 \citep{ta02,jo08} and that the two components (see Section \ref{sec:pre}) will accrete equal mass in the future, each component may increase its mass by only $<$0.04--0.12 M$_{\odot}$ which is close to the stellar/BD boundary.
Given the very uncertain dust opacity and the unpredictable future accretion (i.e. SFE), it is difficult to determine the final mass of the central stars precisely.
However, the very low-mass parent core likely provides material to form only a very low-mass hydrogen-burning binary or a BD binary with one or two substellar objects.

\subsubsection{Mass of central stars derived from orbital motion}
\label{sec:cendis}
%central object mass from orbital model
% Outflow results: dynamical
The wiggling pattern caused by orbital motion in the PV diagram (Figure \ref{fig:pvsma}) gives us the opportunity to calculate the current mass of the central star(s) in IRAS\,16253 using Equation \ref{eq:kep}.
We derive an orbital period $P_{\rm orb}$ of $\sim$3300 yr from the orbital velocity and radius in Table \ref{tab:orbpa}. 
%The binary separation, $a\sim$0\farcs5, can be obtained from $a=R_1+R_2$.
Taking the binary separation $a=R_1+R_2=0\farcs55$ (69\,au), we obtain current stellar masses of 0.026 $M_\odot$ and 0.006 $M_\odot$. The mass ratio comes from $M_1V_{\rm orb,\,1}=M_2V_{\rm orb,\,2}$.
%This result, together with the future accretion (see section \ref{sec:cor}), imply that IRAS\,16253 is forming a low mass star as primary and a BD as companion.
This result, together with the future accretion ($<$0.04--0.12 M$_{\odot}$, see Section \ref{sec:cor}), implies that IRAS\,16253 will form a very low-mass binary system or a BD binary system with at least one BD.
%As mentioned in Section \ref{sec:int}, it supports the idea that a BD can form like a planet. We currently have no clue whether the companion could be ejected later.
Since IRAS\,16253 is located in an isolated environment, our results imply that BDs can form like normal low-mass stars.

Two caveats should be mentioned.
First, the results obtained from the fit to the PV diagrams (Table \ref{tab:orbpa}) depends on the inclination angle.
We derive the total mass of the central stars to be 0.01 M$_{\odot}$ with $\theta_{\rm inc}=10\arcdeg$ and 0.04 M$_{\odot}$ with $\theta_{\rm inc}=30\arcdeg$.
Thus, although the inclination angle affects the estimate of the central mass, it does not change the conclusion that IRAS\,16253 is probably a proto-BD binary system.
Second, the sinusoidal pattern in the PV diagram of the N-S jet (Figure \ref{fig:pvsma}) is not as well fitted by the model as the NE-SW jet, implying that $R_2$ and $V_{\rm orb,\,2}$ are uncertain.
However, this would not affect the orbital period $P_{\rm orb}$ which is obtained from $R_1$ and $V_{\rm orb,\,1}$ and is independent of $R_2$ and $V_{\rm orb,\,2}$; 
we can also replace $\frac{V_{\rm orb}}{R}{\rm t}$ in Equation (\ref{eq:vorb}) by $\frac{2\pi l}{\lambda_{\rm orb}}$ where $\lambda_{\rm orb}$ is the spatial period of the jet due to the orbital motion (see Equation \ref{eq:vpre}), such that the period can be derived with $P_{\rm orb}=\lambda_{\rm orb}/V_{\rm jet}\cos(\alpha')$.
Taking the orbital period, we obtain the total mass of the central stars as a function of $a$ using Equation \ref{eq:kep},
\begin{equation}
M_{\rm tot} = \frac{4\pi^2}{GP_{\rm orb}^2}a^3.
\label{eq:kep2}
\end{equation}
The JVLA continuum map hints that the binary separation is less than 0\farcs5 although we can not exclude the possibility that the companion is too faint to be detected at 43 GHz.  
If we take 0\farcs5 as the upper limit of separation, the total mass of the central stars would be less than 0.032 M$_\odot$.
Given the future accretion $<$0.04--0.12 M$_{\odot}$ (for each component, see Section \ref{sec:cor}), IRAS\,16253 will form a very low-mass binary system which may contain BDs.

\begin{figure}
\includegraphics[scale=.50]{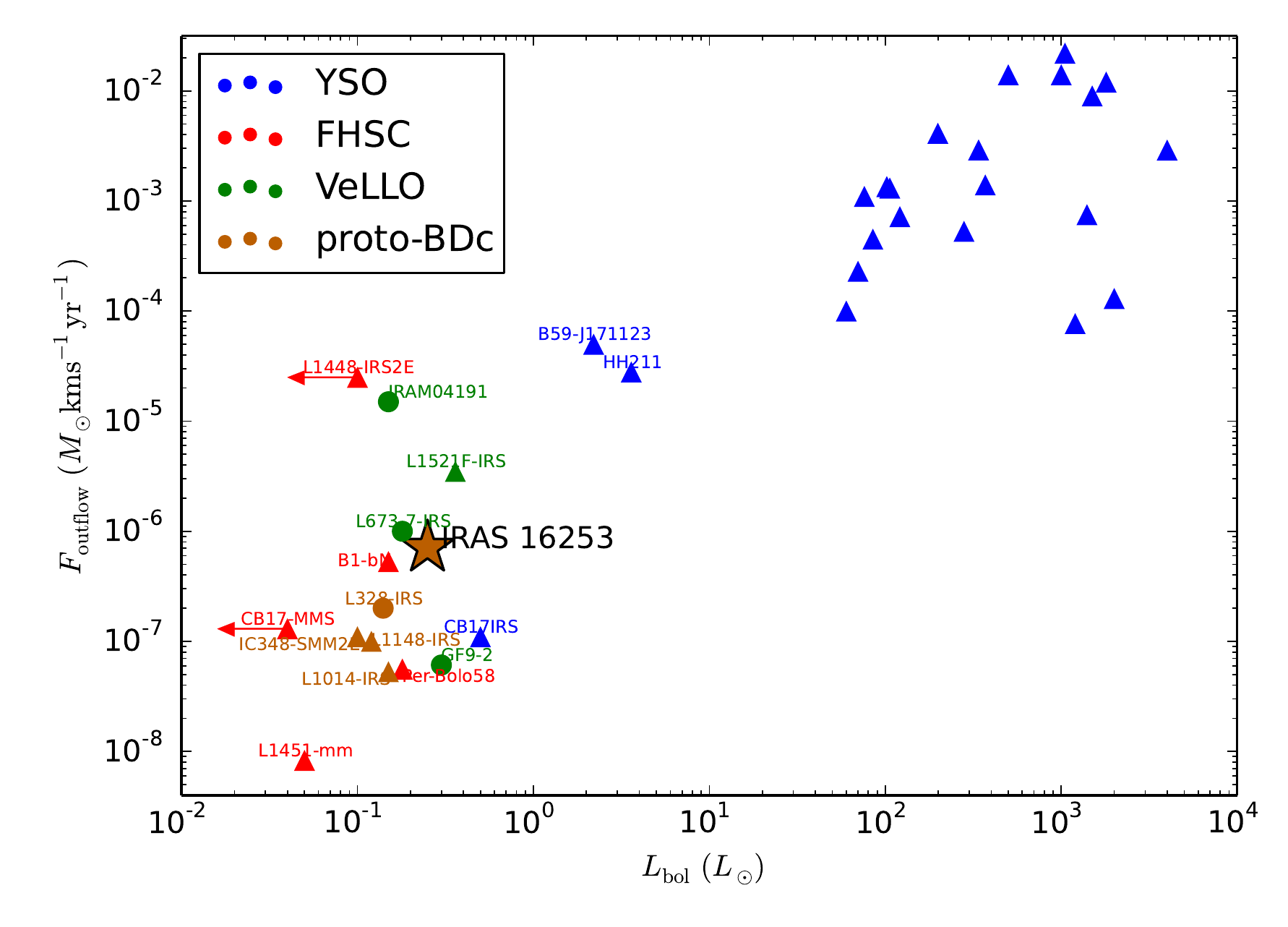}
\caption{Relation between bolometric luminosity and outflow force. 
The figure is reproduced from Figure~7 in \citet{pa14} with IRAS\,16253 (brown star) added for comparison.  
Blue, red, and green points represent YSOs, FHSC, and VeLLOs;
however, we label brown dwarf candidates as brown points which are also VeLLOs. 
Outflow forces measured by interferometers and single dishes are labeled with triangle and circle, respectively. 
The arrows indicate upper limits.
}
\label{fig:lbol}
\end{figure}

\subsubsection{Implication of the outflow force for a proto-BD}
\label{sec:out}
% Outflow result: physical (Compare with Palau IC348)
%add Pho-bo?
%comparison with extremely young and low mass objects
The weak outflow force in IRAS\,16253 also hints at IRAS\,16253 being a proto-BD. A correlation between outflow force and bolometric luminosity $L_{\rm bol}$ has been found and investigated in the past two decades \citep{bo96,wu04}. 
\citet{pa14} compared the proto-BD candidate IC 348-SMM2E with VeLLOs \citep{di07}, First Hydrostatic Cores (FHSCs, Larson 1969), and normal protostars in the $F_{\rm CO}$-$L_{\rm bol}$ plot, in which VeLLOs could be very low-mass protostars or proto-BDs and FHSCs are believed to be the youngest protostars (see Table 6 in Palau et al.\ 2014, and references therein).
We put IRAS\,16253 in this plot (Figure \ref{fig:lbol}), and further highlight four VeLLOs which are considered as proto-BDs (L1014-IRS, Huard et al.\ 2005, Bourke et al.\ 2006; L328, Lee et al.\ 2009, 2013; L1148-IRS, Kauffmann et al.\ 2011; IC 348-SMM2E, Palau et al.\ 2014).
IRAS\,16253 is located between these proto-BD candidates and VeLLOs.
However, IRAS\,16253 is the only source which is likely a binary system among these proto-BD candidates.
Thus, the outflow force and the bolometric luminosity should be considered as upper limits because they are contributed by two objects.

\subsection{Mass accretion}
\subsubsection{Estimate of mass accretion rate from outflow force}
\label{sec:outMacc}
The outflow force is usually used to estimate the mass accretion rate and accretion luminosity $L_{\rm acc}$ in order to understand the nature of VeLLOs \citep{du10a,le13,ta13}.
Here we follow the method of accretion rate estimate in \citet{du10a} and \citet{le13}.
Assuming that the gravitational energy released through a protostellar jet/outflow
is correlated with the mass accretion onto the protostar, 
the mass accretion rate ($\dot{M}_{\rm acc}$) can be represented as
\begin{equation}
\dot{M}_{\rm acc}=\frac{1}{f_{\rm ent}}\frac{\dot{M}_{\rm acc}}{\dot{M}_W}\frac{1}{V_W}F_{\rm out},
\label{eq:macc}
\end{equation}
where $f_{\rm ent}$ is the entrainment efficiency, $\dot{M}_W$ is the mass-loss rate, and $V_W$ is the jet velocity \citep{bo96}.
We adopt $f_{\rm ent}=0.1$, $\dot{M}_W/\dot{M}_{\rm acc}=0.1$, and $V_W=150$ km s$^{-1}$ \citep{an99,du10a}, and obtain a mass accretion rate of
$4.7\times 10^{-7} M_\odot$ yr$^{-1}$ for IRAS\,16253.
Assuming 10\% mass loss through jet or wind \citep{bo96}, we derive the accreted mass during the Class 0 lifetime of $\sim$0.16 Myr  \citep{an00,ev09} to be $\sim$0.068 $M_\odot$.
If a star accretes half of its mass at the Class 0 stage, IRAS\,16253 could only reach a terminal total mass of $\sim$0.14 $M_\odot$ (including two components).
This result may be considered as an upper limit because the mass accretion rate is believed to decrease as the core evolves \citep{bo96} and
 IRAS\,16253 with a bolometric temperature of 27 K is likely a very young Class 0 object \citep{hs15}.
This result is consistent with the conclusions in Sections \ref{sec:cor} and \ref{sec:cendis}.
We note that the jet velocity of proto-BDs was recently found to be 50--100 km s$^{-1}$ \citep{mo15}, which would yield an increase in mass accretion rate and terminal total mass by a factor of 1.5--3, i.e. a mass of 0.21--0.42 $M_\odot$. However, given the parent core mass of $<0.82 M_\odot$, the higher value would imply a SFE of 0.5, which is unlikely.

Because the accretion luminosity is one of the crucial parameters for understanding the low luminosity of VeLLOs, we calculate the accretion luminosity $L_{\rm acc}$ using
\begin{equation}
L_{\rm acc}=\frac{GM_{\rm acc}\dot{M}_{\rm acc}}{R}
\label{eq:lacc}
\end{equation}
where $G$ is the gravitational constant, $M_{\rm acc}$ is the accreted mass ($\dot{M}_{\rm acc}\times\tau_{\rm dyn}$), and $R$ is the protostellar radius ($3R_\odot$).
The resulting accretion luminosity is $\sim$0.03 $L_\odot$. Given the uncertainties on $L_{\rm acc}$, this value is comparable to the internal luminosity of 0.08--0.09$L_\odot$ \citep{du08}.

\subsubsection{Probe of episodic accretion}

%Episodic accretion and luminosity problem
Episodic accretion has been proposed \citep{du10b,du12,jo15} to solve the long-standing luminosity problem namely that the observed bolometric luminosities as well as mass accretion rates in protostars are much lower than expected \citep{ke90}.
The discovery of VeLLOs has further exacerbated the luminosity problem and VeLLOs have become important for studying it.
In episodic accretion models, a protostellar system is at a quiescent accretion stage for most of the time and accretion bursts occasionally occur to deliver material onto the central protostar \citep{ke95,le07,du10b,du12,jo15}. This behavior leads to protostars having a low luminosity for most of the time but still accreting sufficient mass.

\citet{jo15} proposed that a recent accretion burst can be probed by comparing the observed extent of C$^{18}$O emission with the predicted extent given by the current source luminosity.
An accretion burst would enhance the luminosity (i.e., accretion luminosity $L_{\rm acc}$) which heats the parent core and shift the CO freeze-out/sublimation boundary to a larger radius.
After a burst, CO would take a long time to refreeze-out onto the dust surfaces ($10^3-10^4$ yr, Visser et al. 2015) while the accretion luminosity has decayed.
The C$^{18}$O extent in IRAS\,16253 with a radius of $\sim$210 au ($\sqrt{4\farcs2\times2\farcs8}/2=1\farcs7$, see Section \ref{sec:cen}) is significantly larger than the expected radius $\sim$100 au considering the bolometric luminosity of $0.25\, L_{\odot}$ (see Figure 4 in J\o rgensen et al.\ 2015); 
the expected radius is where the temperature reaches the CO sublimation region (20--30 K) produced by the heating from the central object.
The luminosity needed to reproduce the C$^{18}$O extent is a factor of $\sim$4 higher than the current bolometric luminosity of IRAS\,16253.
Because the C$^{18}$O emission shows no clear correlation with the jets on large scales (Figure \ref{fig:sma}), we suggest that the C$^{18}$O emission is associated with the inner dense region of the envelope as the sources in \citet{jo15}.
As a result, IRAS\,16253 is likely at a post-burst stage (or has undergone a recent accretion burst) in the episodic accretion process.

%Our models
We set a model with one accretion burst that occurred during the jet dynamical time ($1.3\times10^4$ yr) in order to examine whether we can detect the occurrence of a burst based on the jet locus. In this model, we assume that a burst occurred at -6000 yr and the mass linearly increased within 1000 yr (end at -5000 yr).
Our model predicts a transition in the jet locus due to the accretion burst: the trajectory changes abruptly in Figure \ref{fig:pvmod}c, with rapid variations in amplitude and period.
Because our model simply considers Kepler's law, the path at the transition phase 
may not be exactly true.
Nevertheless, we find that the wiggling patterns are very different before and after an accretion burst.
Therefore, we suggest that with additional high spatial and spectral resolution data covering the whole jet, modeling the jet wiggling pattern could help identifying the accretion process (episodic or continuous) and assessing whether accretion bursts have occurred.

\section{Summary}
We identified a proto-brown dwarf (proto-BD) binary system candidate through its dynamics for the first time.
Based on multi-transition CO (2--1/6--5/7--6) observations from single dishes (IRAM 30 m and APEX) and an interferometer (SMA),
we studied the dynamical and physical properties of the protostellar jets driven by IRAS\,16253 in detail.
The ``S-shaped'' jet detected in H$_2$ and CO (6--5) suggests that IRAS\,16253 hosts a close binary system.
The SMA CO (2--1) data further reveal the jet wiggling caused by the orbital motion in the PV diagram, which allows us to probe the dynamics of the central binary system.
Furthermore, we use the multi-transition CO data to derive the gas temperature and column density of the outflows.
As a result, we obtain the outflow mass, momentum, and force of IRAS\,16253.
We also find that the small-scale emission of the mid-J CO transitions matches the H$_2$ emission and probes the shocked gas while the low-J CO transition does not.
Here we summarize the properties of (1) the proto-BD binary candidate IRAS\,16253 and (2) its jets/outflows:

\begin{enumerate}
\item The proto-brown dwarf binary candidate IRAS\,16253:
\begin{enumerate}
\item Using the jet-wiggle model, we derive the current total mass of the binary system to be 0.032$~\pm~0.003~M_{\odot}$. The low parent core mass ($\leqslant0.8M_\odot$) further suggests that 
IRAS\,16253 will form a very low-mass binary or a BD binary in the future.

\item The low outflow force implies a very low mass accretion rate in IRAS\,16253 in the main accretion phase, suggesting that it will probably form substellar objects.
The outflow force versus bolometric luminosity plot also hints at IRAS\,16253 being a proto-BD binary candidate.
\item The extended C$^{18}$O emission, together with the N$_2$H$^+$ depletion, implies that IRAS\,16253 has experienced an accretion burst in the last $\sim$10$^4$ yr which is about the dynamical age of the protostellar jet.

\item Since IRAS\,16253 is located in an isolated environment, our result supports a scenario in which BDs form through fragmentation and collapse like normal hydrogen-burning stars.

\end{enumerate}
\item The jets/outflows driven by IRAS\,16253:
\begin{enumerate}

\item The CO (2--1) emission from IRAM 30~m primarily traces the entrained gas and/or outflow cavity. The CO (6--5) emission can be decomposed into two components: (1) the large-scale outflow cavity wall presumably heated by UV photons from a jet-driven bow shock and (2) the high velocity shocked gas heated by the interaction of the jet and the envelope.
\item The outflow energies and the forces derived from the small-scale SMA CO (2--1) jets are comparable with that of the large-scale IRAM CO (2--1) outflows, suggesting that the entrained gas can be driven by the collimated jets.
\item We modeled the jet wiggling pattern caused by the orbital motion in the position-velocity diagram for a protostellar jet driven by a binary system.
The model could be used to trace the history of the binary formation including the orbital evolution and accretion process.
\end{enumerate}
\end{enumerate}
\acknowledgments
The authors thank Dr. J. J. Tobin for providing the CARMA N$_2$H$^+$ (1--0) data for Figure \ref{fig:sma}. 
We are grateful to Dr. A. Palau for giving the data points in Figure \ref{fig:lbol}.
We would like to thank Dr. T. Stanke for providing us the JCMT CO (3--2) data for comparison.
The authors thank the referee for the comments that improved this paper.
The authors acknowledge the staff at APEX, IRAM 30~m, SMA, and JVLA for assistance with operations.
The Submillimeter Array is a joint project between the Smithsonian Astrophysical Observatory and the Academia Sinica Institute of Astronomy and Astrophysics and is funded by the Smithsonian Institution and the Academia Sinica.
T.H.H and S.P.L. acknowledge support from the Ministry of Science and
Technology (MOST) of Taiwan with Grants MOST 102-2119-M-007-004- MY3.
T.H.H appreciates the grant from MOST with the PhD exchange student award (MOST 103-2917-I-007-005) and also thanks the MPIfR for supporting him as a visiting astronomer in Bonn, Germany.

\clearpage
\appendix
\section{Original maps and channel maps}

\setcounter{figure}{0}
\setcounter{table}{0}
\renewcommand{\thefigure}{A\arabic{figure}}
To study the molecular outflows, we remove the large-scale CO emissions in our single-dish observations using a multiresolution analysis (see Section \ref{sec:sca}, Belloche et al.\ 2011) because they are mostly contributed from the molecular cloud and core.
This analysis decomposes the intensity maps into summation maps (small-scale) and smooth maps (large-scale) which correspond to the outflow and cloud emissions, respectively.
We apply this analysis to our CO (2--1)/(6--5)/(7--6) channel maps and all the analyses done in this paper are based on the small-scale outflow maps;
the outflow properties are derived from the step 4 CO (2--1), step 5 CO (6--5), and step 4 CO (7-6) summation maps which correspond to the structures at scales less than 17 pixels, 31 pixels, and 17 pixels with a pixel size of 5\arcsec.
We show the integrated intensity maps of the input map and the output maps (summation maps and smooth maps) in Figure~\ref{fig:sca}.
The channel maps of summation maps and smooth maps are shown in Figures \ref{fig:IRAMch1} to \ref{fig:APEXch1bsmo}.

\begin{figure}[b]
\includegraphics[scale=0.55]{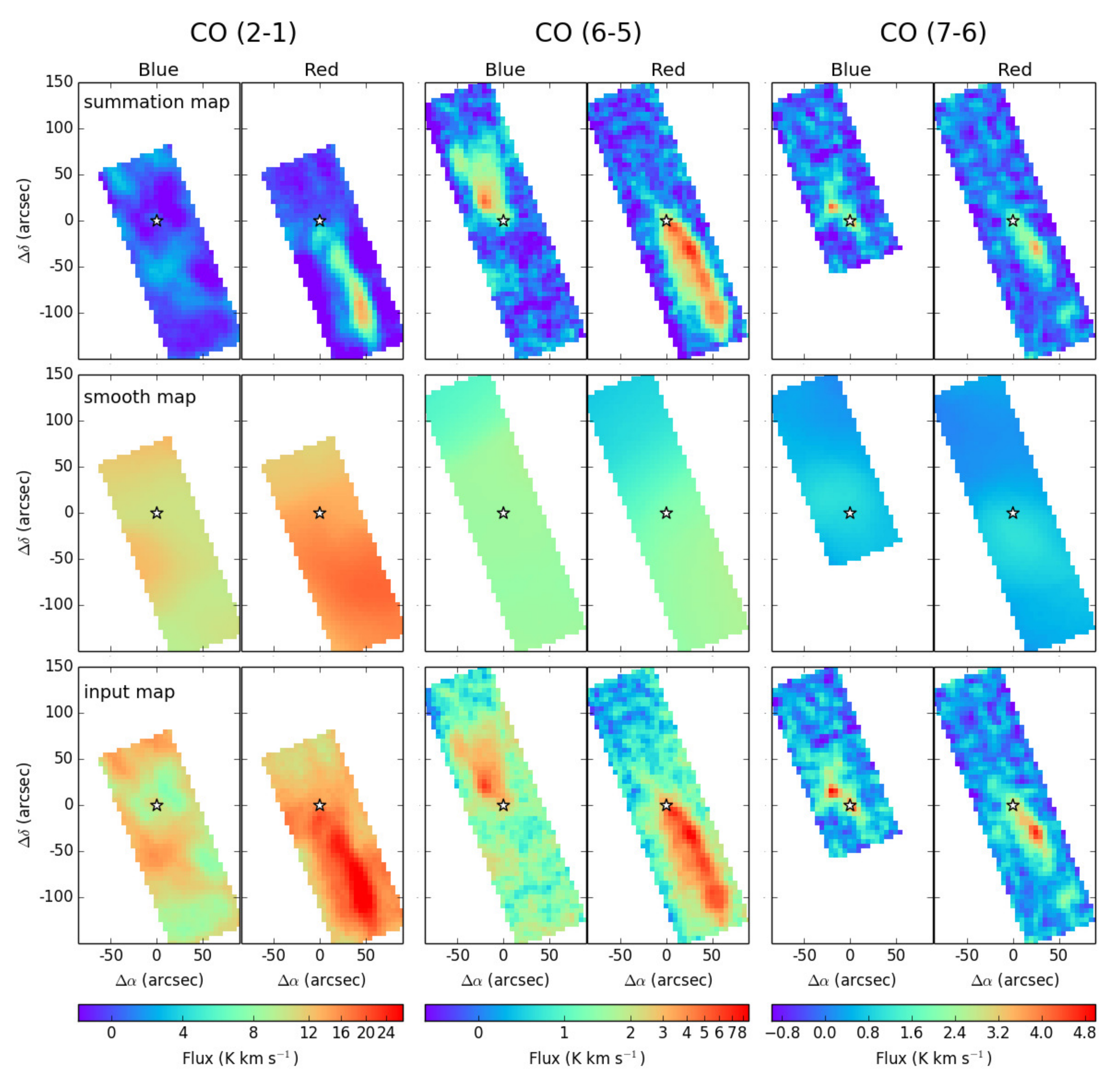}
\caption{Comparison of integrated intensity maps before and after removing large-scale structures. 
The left, middle, and right panels represent CO (2--1), (6--5), and (7--6) maps, respectively, and the top, middle, and bottom panels represent the summation, smooth, and input maps, respectively (see section \ref{sec:sca}). The blue-shifted (right) and red-shifted (left) integrated maps have the same velocity ranges as in Figure~\ref{fig:sin}.
}
\label{fig:sca}
\end{figure}

\begin{figure*}
\includegraphics[scale=.42]{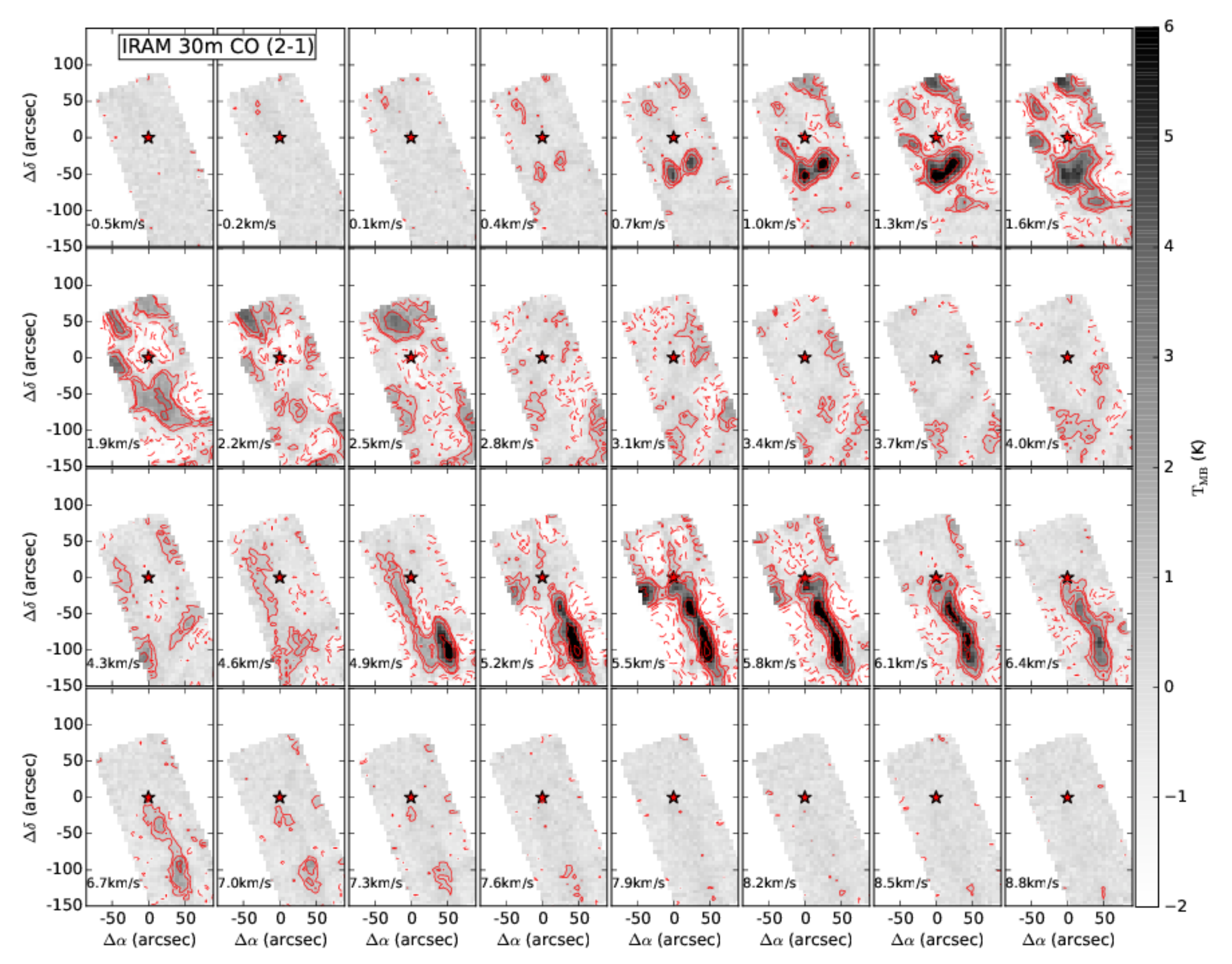}
\caption{IRAM CO (2--1) channel maps after filtering the large-scale emission. These channel maps were processed through the multiresolution analysis of \citet{be11} and represent the step 4 summation maps. The contour levels are -10, -5, -3, 3, 5, 10, 20, and 30$\sigma$ with a rms noise level $\sigma =$ 0.26 K. The red star indicates the position of the infrared source.
}
\label{fig:IRAMch1}
\end{figure*}

\begin{figure*}
\includegraphics[scale=.42]{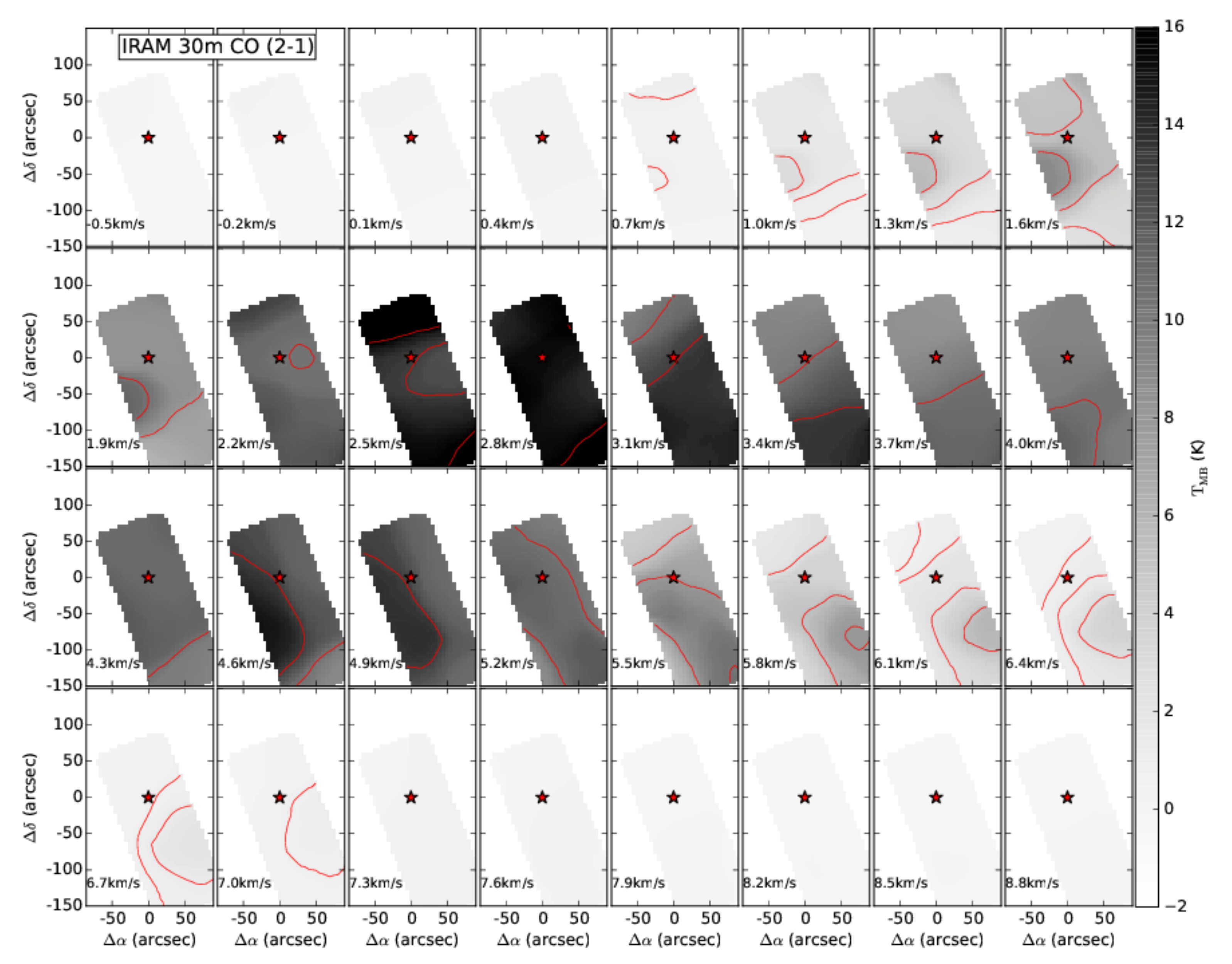}
\caption{IRAM CO (2--1) channel maps of large-scale emission which is filtered in Figure \ref{fig:IRAMch1}. The contour levels are the same as those of Figure \ref{fig:IRAMch1}.
}
\label{fig:IRAMch1smo}
\end{figure*}

\begin{figure*}
\includegraphics[scale=.42]{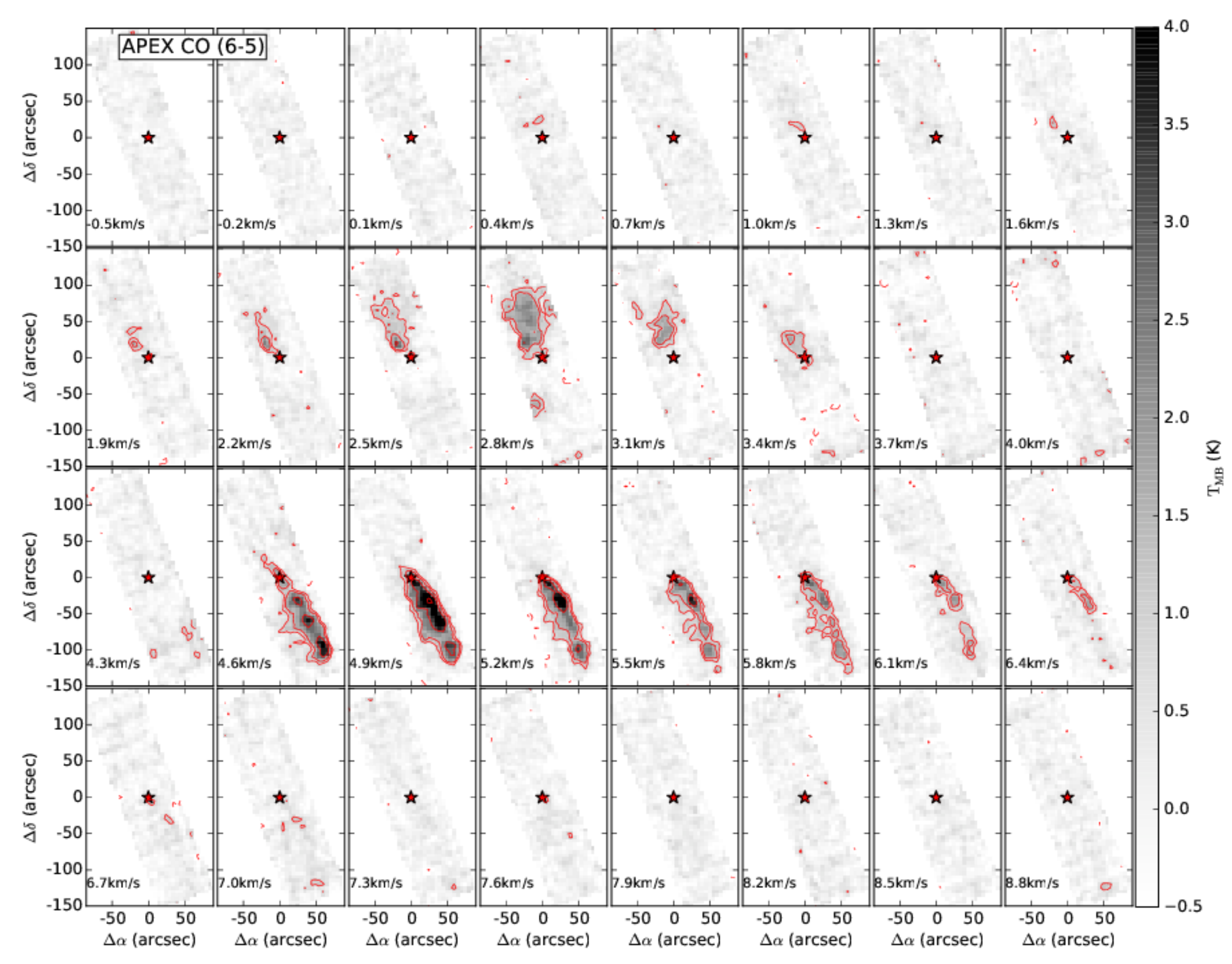}
\caption{APEX CO (6--5) channel maps after filtering the large-scale emission. These channel maps were processed through the multiresolution analysis of \citet{be11} and represent the step 5 summation maps. The contour levels are -10, -5, -3, 3, 5, 10, and 20$\sigma$ with a rms noise level $\sigma =$ 0.28 K.
}
\label{fig:APEXch1}
\end{figure*}

\begin{figure*}
\includegraphics[scale=.42]{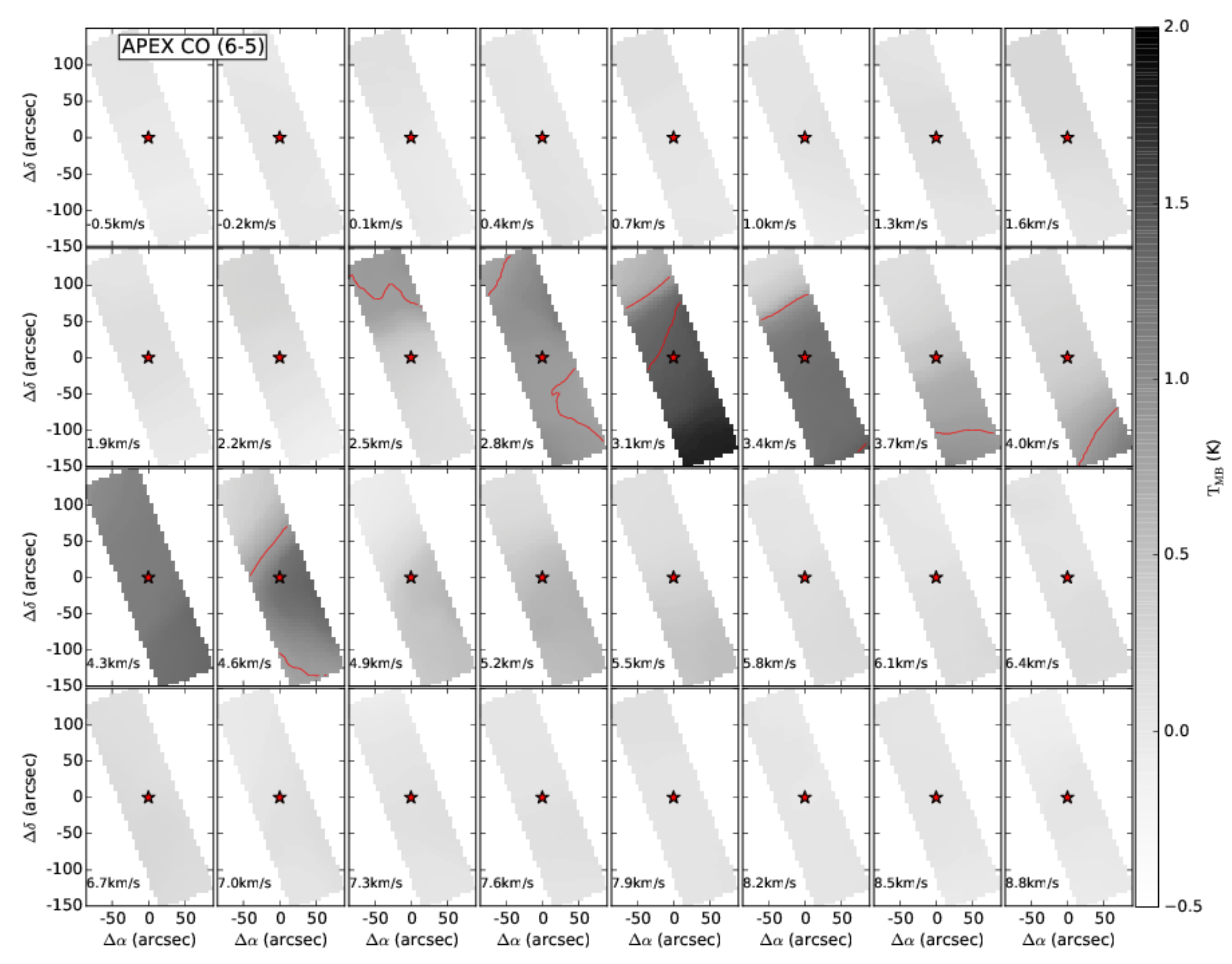}
\caption{APEX CO (6--5) channel maps of large-scale emission which is filtered in Figure \ref{fig:APEXch1}. The contour levels are the same as those of Figure \ref{fig:APEXch1}.
}
\label{fig:APEXch1smo}
\end{figure*}

\begin{figure*}
\includegraphics[scale=.42]{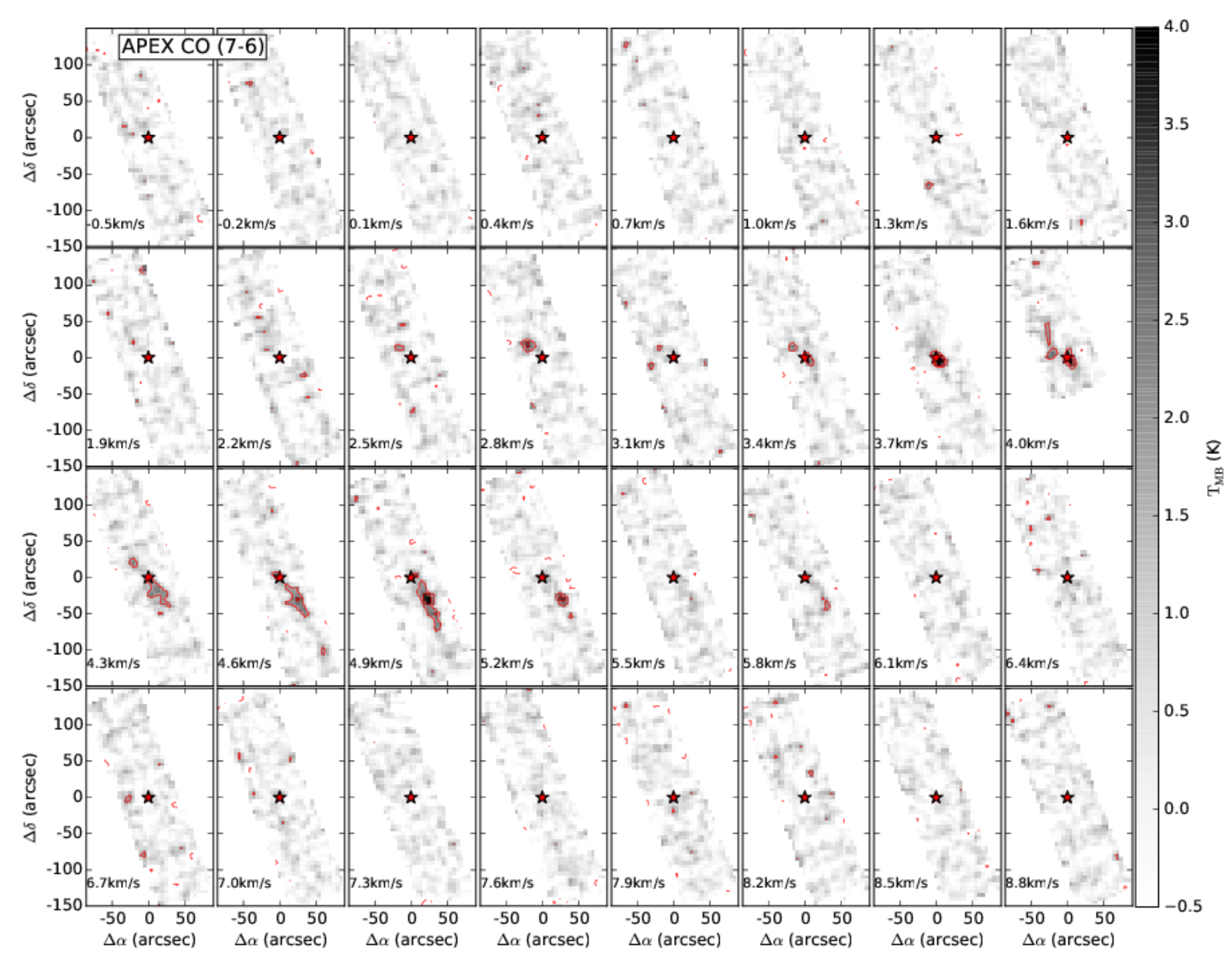}
\caption{APEX CO (7--6) channel maps after filtering the large-scale emission. These channel maps were processed through the multiresolution analysis of \citet{be11} and represent the step 4 summation maps. The contour levels are 3 and 5$\sigma$ with a rms noise level $\sigma =$ 0.57 K.
}
\label{fig:APEXch1b}
\end{figure*}

\begin{figure*}
\includegraphics[scale=.42]{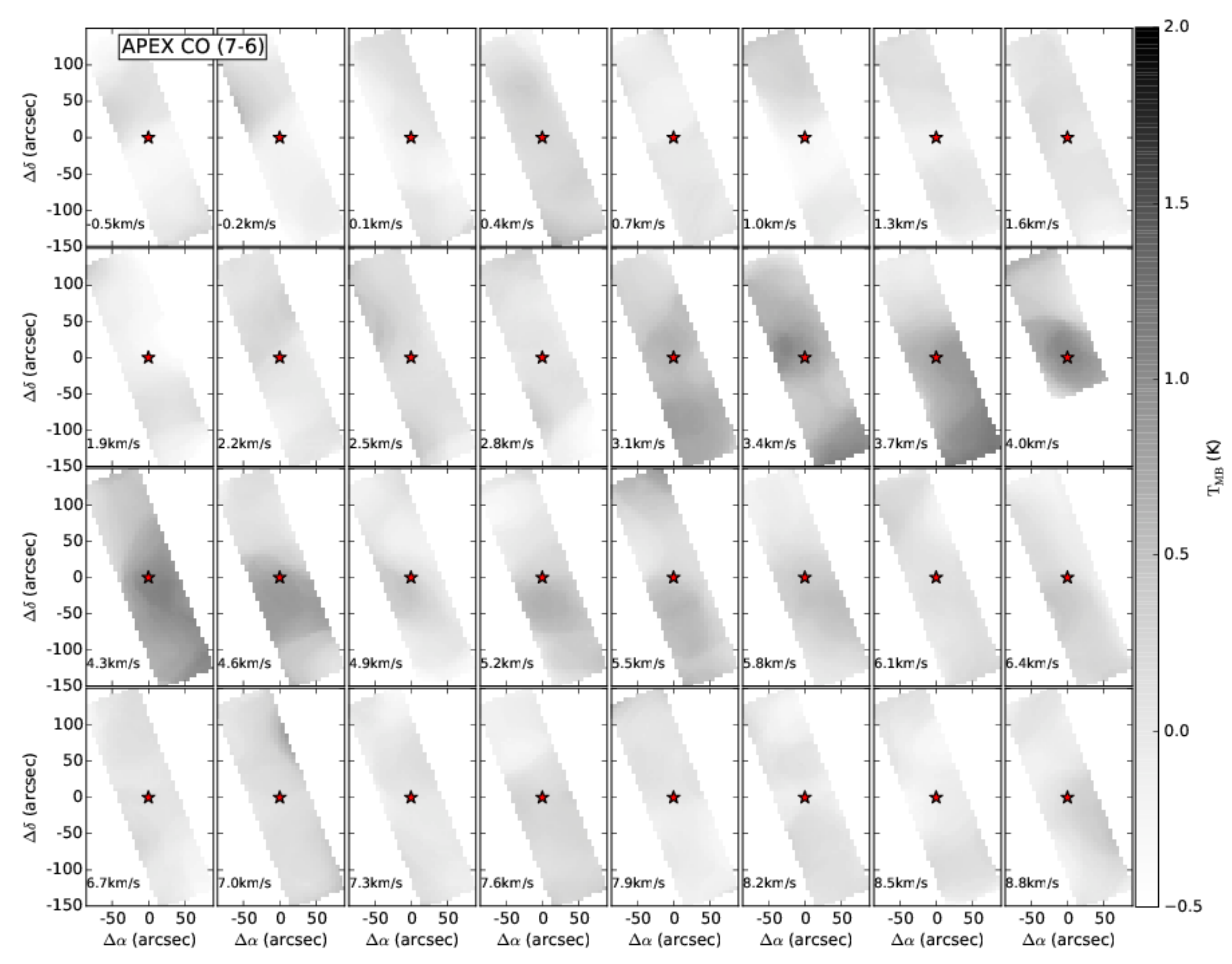}
\caption{APEX CO (7--6) channel maps of large-scale emission which is filtered in Figure \ref{fig:APEXch1b}. The contour levels are the same as those of Figure \ref{fig:APEXch1b}.
}
\label{fig:APEXch1bsmo}
\end{figure*}

\end{document}